\documentclass[10pt,letterpaper]{article}
\usepackage{enumitem}

\usepackage[top=0.85in,left=2.75in,footskip=0.75in]{geometry}

\usepackage{amsmath,amssymb}

\usepackage{changepage}

\usepackage[utf8x]{inputenc}

\usepackage{textcomp,marvosym}

\usepackage{cite}

\usepackage{nameref,hyperref}

\usepackage[right]{lineno}

\usepackage{microtype}
\DisableLigatures[f]{encoding = *, family = * }

\usepackage[table]{xcolor}

\usepackage{array}

\newcolumntype{+}{!{\vrule width 2pt}}

\newlength\savedwidth

\raggedright
\usepackage[aboveskip=1pt,labelfont=bf,labelsep=period,justification=raggedright,singlelinecheck=off]{caption}

\makeatletter
\renewcommand{\@biblabel}[1]{\quad#1.}
\makeatother

\usepackage{lastpage,fancyhdr,graphicx}
\usepackage{epstopdf}
\fancyheadoffset[L]{2.25in}
\fancyfootoffset[L]{2.25in}
\begin{document}
\vspace*{0.2in}

\begin{flushleft}
{\Large
\textbf\newline{Influence of confinement on the spreading\\of bacterial populations} 
}
\newline
\\
Daniel B. Amchin\textsuperscript{1}, Jenna A. Ott\textsuperscript{1}, Tapomoy Bhattacharjee\textsuperscript{2}, and Sujit S. Datta\textsuperscript{1*}\\

\bigskip
\textbf{1} Department of Chemical and Biological Engineering, Princeton University, Princeton, New Jersey, United States
\\
\textbf{2} 
Andlinger Center for Energy and the Environment, Princeton University, Princeton, New Jersey, United States
\\
\bigskip

%
%





* Correspondence to ssdatta@princeton.edu

\end{flushleft}
\section*{Abstract}
 The spreading of bacterial populations is central to processes in agriculture, the environment, and medicine. However, existing models of spreading typically focus on cells in unconfined settings --- despite the fact that many bacteria inhabit complex and crowded environments, such as soils, sediments, and biological tissues/gels, in which solid obstacles confine the cells and thereby strongly regulate population spreading. Here, we develop an extended version of the classic Keller-Segel model of bacterial spreading that incorporates the influence of confinement in promoting both cell-solid and cell-cell collisions. Numerical simulations of this extended model demonstrate how confinement fundamentally alters the dynamics and morphology of spreading bacterial populations, in good agreement with recent experimental results. In particular, with increasing confinement, we find that cell-cell collisions increasingly hinder the initial formation and the long-time propagation speed of chemotactic pulses. Moreover, also with increasing confinement, we find that cellular growth and division plays an increasingly dominant role in driving population spreading---eventually leading to a transition from chemotactic spreading to growth-driven spreading \textit{via} a slower, jammed front. This work thus provides a theoretical foundation for further investigations of the influence of confinement on bacterial spreading. More broadly, these results help to provide a framework to predict and control the dynamics of bacterial populations in complex and crowded environments.

\section*{Author summary}
The spreading of bacteria through their environments critically impacts our everyday lives; it can be harmful, underlying the progression of infections and spoilage of foods, or can be beneficial, enabling the delivery of therapeutics, sustaining plant growth, and remediating polluted terrain. In all these cases, bacteria typically inhabit crowded environments, such as soils, sediments, and biological tissues/gels, in which solid obstacles confine the cells and regulate their spreading. However, existing models of spreading typically focus on cells in unconfined settings, and thus are frequently not applicable to cells in more complex environments. Here, we address this gap in knowledge by extending the classic Keller-Segel model of bacterial spreading to incorporate the influence of confinement. Through mathematical analysis and numerical simulations of this extended model, we show how confinement fundamentally alters the dynamics and morphology of spreading bacterial populations --- in particular, driving a transition from chemotactic spreading of motile cells to growth-driven spreading \textit{via} a slower, jammed front. These results provide a foundation for further investigations of the influence of confinement on bacterial spreading, both by yielding testable predictions for future experiments, and by providing guidelines to predict and control the dynamics of bacterial populations in complex and crowded environments.
\linenumbers

\section{Introduction}\label{Intro}
The ability of bacterial populations to spread through their surroundings plays a pivotal role in our everyday lives. It can be harmful, underlying the progression of infection in the body \cite{balzan,chaban,pnas,harman,ribet,siitonen,lux,oneil} and the spoilage of foods \cite{Gill1977,Shirai2017}. In other cases, it can be beneficial, enabling bacteria to deliver drugs to hard-to-reach spots in the body \cite{thornlow,toley}, move toward and protect plant roots in soil \cite{dechesne,souza,turnbull,watt,babalola}, and degrade environmental contaminants \cite{roseanne18,Adadevoh:2016,ford07,wang08,reddy}. Therefore, the development of accurate models of spreading is critically important for the prediction and control of bacterial populations in medicine, food, agriculture, and the environment.


One common way in which many bacteria spread is through active motility using, for example, flagellar propulsion in liquids or pili-mediated propulsion on solids. Populations of cells can thereby direct their spreading \textit{via} chemotaxis: as the cells continually consume a surrounding chemoattractant, such as a nutrient or oxygen, they collectively generate a local gradient that they, in turn, bias their motion along. This biased motion leads to the spectacular formation of a coherent pulse of cells that continually propagates, enabling populations to escape from harmful conditions and colonize new terrain \cite{adler1966effect,adler1966science,saragosti2011directional,fu2018spatial,cremer2019chemotaxis,bhattacharjee2021chemotactic}. Building on the seminal work of Keller and Segel in 1971, continuum-scale models have been developed that can successfully capture the key features of this chemotactic spreading in bulk liquids \cite{keller1971traveling,odell1976traveling,keller1975necessary,lauffenburger1991quantitative,fu2018spatial,cremer2019chemotaxis,seyrich2019traveling}; hence, such models form a cornerstone of theoretical studies of emergent phenomena and collective behavior in biology.

However, while such models and the lab studies that they describe typically focus on cells in unconfined environments, bacteria typically inhabit more crowded settings -- such as porous gels and tissues in the body, micro- and meso-porous foods, and soils, sediments, and subsurface formations in the environment -- in which a solid matrix obstructs and confines the motion of cells. Depending on the degree of confinement, bacterial populations may still be able to spread \textit{via} chemotaxis \cite{croze2011migration,bhattacharjee2021chemotactic}, but with two notable differences, as revealed by recent experiments. First, as confinement increases, collisions with the solid matrix increasingly impede the migration of individual cells \cite{cisneros2006reversal,drescher2011fluid,tapasoftmatter,051bhattacharjeenatcomm}. Second, also as confinement increases, the amount of free space available for cells to move through decreases, promoting cell-cell collisions that further impede cellular migration \cite{bhattacharjee2021chemotactic}. Indeed, if the cells are sufficiently confined and their local density is sufficiently high, cell-cell collisions dominate and they ultimately become jammed, unable to self-propel at all \cite{dell2018growing,volfson2008biomechanical} --- abolishing chemotaxis entirely. In this case, the population instead spreads solely through cellular proliferation, referred hereafter as `growth' for brevity, in which metabolically-active cells grow, divide, and push each other to new locations \cite{farrell2013mechanically}. Hence, confinement -- which is an inherent feature of many bacterial habitats -- is a strong regulator of population spreading.

Unfortunately, despite its widespread use, the classic Keller-Segel model treats cells as being non-contacting, does not explicitly incorporate confinement, and does not consider population spreading \textit{via} growth. Previous work took a useful first step toward extending this model by modifying the underlying transport parameters to incorporate the influence of cell-solid collisions \cite{croze2011migration}. Nevertheless, due to the limited understanding of single-cell motility in confinement at the time, this approach necessarily relied on \textit{ad hoc} approximations; moreover, it did not incorporate cell-cell collisions or jamming. As a result, there remains a need for models that can more accurately describe the spreading of bacteria in crowded and highly-confining environments.  

Here, we present an extended version of the classic Keller-Segel model that incorporates the influence of confinement on bacterial spreading through both cell-solid and cell-cell collisions, motivated by recent experimental observations. We identify key dimensionless parameters emerging from this extended model that describe bacterial spreading. Furthermore, by numerically solving the model, we show how confinement fundamentally alters the dynamics and morphology of spreading bacterial populations. In particular, with increasing confinement, we find that cell-cell collisions increasingly hinder the initial formation and the long-time propagation speed of chemotactic pulses. Moreover, also with increasing confinement, growth plays an increasingly dominant role in driving population spreading compared to cellular motility---eventually leading to a transition from chemotactic spreading to growth-driven spreading \textit{via} a slower, jammed front. Thus, our work provides a foundation for future investigations of the influence of confinement, and yields quantitative principles that could guide the prediction and control of bacterial spreading in crowded and complex environments.

\section{Methods}\label{Methods}
\subsection{Classic Keller-Segel model}\label{Methods:ClassicKSModel}
\noindent We first describe chemotactic migration using the classic one-dimensional Keller–Segel model, which does not incorporate the influence of confinement, but can successfully capture the key features of experiments on dilute populations of bacteria in bulk liquid \cite{fu2018spatial,saragosti2011directional,cremer2019chemotaxis,croze2011migration,keller1975necessary,keller1971traveling,odell1976traveling,lauffenburger1991quantitative,seyrich2019traveling}. As is conventionally done \cite{lauffenburger1991quantitative,keller1971traveling,adler1966science}, and to directly connect the model to many experiments \cite{bhattacharjee2021chemotactic,croze2011migration,adler1966science,fu2018spatial}, we consider a sole nutrient that also acts as the chemoattractant, represented by the continuum variable $c(x,t)$, where $x$ is the position coordinate and $t$ is time. The number density of bacteria, in turn, is given by the continuum variable $b(x,t)$. Furthermore, given the experimental conditions, we assume that the cells do not excrete their own chemoattractant or other diffusible signals, as is sometimes the case in low-nutrient conditions and for specific strains. Recent extensions of this model have also considered the case in which nutrient and attractant are separate chemical species, which leads to fundamentally different behavior that would be interesting to explore using our framework in future work \cite{cremer2019chemotaxis,narla2021traveling}.

As the nutrient diffuses through space with thermal diffusivity $D_c$, it is consumed by the cells at a rate $b\kappa g(c)$; here, $\kappa$ is the maximum consumption rate per cell and the Monod function $g(c)\equiv {c}/({c + c_{\rm{char}}})$, with the characteristic concentration $c_{\rm{char}}$, quantifies the reduction in consumption rate when nutrient is sparse \cite{croze2011migration,monod1949growth,cremer2019chemotaxis,woodward1995spatio,shehata1971effect,Schellenberg1977,cremer2016effect}. Therefore, the nutrient dynamics are given by 
\begin{equation}\label{eqnKellerSegelc}
\frac{\partial c}{\partial t} = D_{\rm{c}} \nabla^2 c \ - \ b\kappa g(c).
\end{equation}

The bacterial dynamics have two contributions: a motility-driven flux $\vec{J}_m$ and cellular proliferation. The flux arises from the combination of the undirected spreading of cells, a diffusive process with an active diffusivity $D_{b0}$ \cite{bergecoli}, and directed chemotaxis with a drift velocity $\vec{v}_{c}\equiv\chi_{0} \nabla f(c)$ that quantifies the ability of the bacteria to logarithmically respond to the local nutrient gradient \cite{keller1971traveling,odell1976traveling,keller1975necessary}. The well-established function $f(c)\equiv\log\left[({1+c/c_{-}})/({1+c/c_{+}})\right]$ quantifies the ability of the cells to sense nutrient with characteristic bounds $c_{-}$ and $c_{+}$ \cite{cremer2019chemotaxis,sourjik2012responding,shimizu2010modular,tu2008modeling,Kalinin2009,shoval2010fold,lazova2011response,celani2011molecular,fu2018spatial,dufour,yang2015relation}, while the chemotactic coefficient $\chi_{0}$ quantifies the ability of the cells to bias their motion in response to a sensed nutrient gradient. Therefore, the motility-driven flux $\vec{J}_m=-D_{{b0}} \nabla b+b\vec{v}_c$. Proliferation, on the other hand, is given by $b\gamma g(c)$, where $\gamma$ is the maximal growth rate per cell and $g(c)$ reflects the reduction in growth rate when nutrient is sparse --- circumventing the need to introduce an \textit{ad hoc} ``carrying capacity" of a logistically-growing population, as is sometimes done. Therefore, the bacterial dynamics are given by 
\begin{equation}\label{eqnKellerSegelb}
\frac{\partial b}{\partial t} = \underbrace{D_{{b0}} \nabla^{2} b  -  \nabla \cdot (bv_{\rm{c}})}_{-\nabla\cdot\vec{J}_{m}} \ + \ b\gamma g(c).
\end{equation}
\noindent Together, Equations \ref{eqnKellerSegelc}-\ref{eqnKellerSegelb} represent the classic Keller-Segel model that describes the coupled dynamics of nutrient and bacteria. In particular, they successfully capture the key features of chemotactic migration in unconfined liquid, in which cells collectively generate a local gradient of nutrient that they in turn bias their motion along---leading to the formation of a coherent pulse of bacteria that continually propagates, sustained by its continued consumption of the surrounding attractant \cite{adler1966science,adler1966effect,fu2018spatial,saragosti2011directional}.

\subsection{Characteristic dimensionless parameters}\label{Methods:DimlessParam}
\noindent Non-dimensionalizing the Keller-Segel equations yields useful dimensionless parameters for characterizing population spreading. We rescale $\{c,b,t,x\}$ by the characteristic quantities $\{c_{\infty},b_0,t_{c,0},\zeta\}$, where $c_{\infty}$ is the initial nutrient concentration taken to be constant everywhere, $b_{0}$ is the maximal initial cell density, $t_{c,0}\equiv c_\infty/\left(b_{0} \kappa\right)$ is a characteristic time scale of nutrient consumption, and $\zeta_{0}\equiv\sqrt{D_{{b0}} t_{c,0}}$ is the characteristic extent of cellular diffusion over the duration $t_{c,0}$. This process yields the non-dimensional equations 
\begin{equation}\label{dimlessEqnKellerSegelc}
\frac{\partial \tilde{c}}{\partial \tilde{t}} = \delta_0 \tilde{\nabla}^2 \tilde{c} \ - \tilde{b} \tilde{g}
\end{equation}
\begin{equation}\label{dimlessEqnKellerSegelb}
\frac{\partial \tilde{b}}{\partial \tilde{t}} = \tilde{\nabla}^{2} \tilde{b} \ - \alpha_0 \tilde{\nabla} \cdot (\tilde{b}\tilde{\nabla}{ \tilde{f}}) \ + \beta_0\tilde{b} \tilde{g},\end{equation}
where the tildes indicate non-dimensionalized variables. Three dimensionless parameters emerge: 
\begin{itemize}
\item The diffusion parameter $\delta_0\equiv D_c/D_{b0}$ compares the thermal diffusion of nutrient to the active diffusion of bacteria. 
When $\delta_0 \ll 1$, 
variations in nutrient are localized to the leading edge of the bacterial population, whereas when $\delta_0 \gg 1$, nutrient levels vary over large spatial extents.

\item The directedness parameter $\alpha_0 \equiv \chi_0/D_{b0}$ compares the influence of chemotaxis to active diffusion in driving cellular spreading. When $\alpha_0 \ll 1$, diffusion dominates and cells do not appreciably direct their motion in response to a nutrient gradient, whereas when $\alpha_0 \gg 1$, motile cells strongly direct their motion in response to a gradient. 

\item The yield parameter $\beta_0 \equiv {\gamma}/({b_{0} \kappa /c_\infty})$ compares the rates of cell growth and nutrient consumption, $\gamma$ and $t_{c,0}^{-1}$, respectively. It therefore quantifies the yield of new cells produced as a population consumes nutrient. 
When $\beta_0 \ll 1$, nutrient consumption is much faster than proliferation, whereas when $\beta_0 \gg 1$, many new cells are produced in the time required to consume the available nutrient. 
\end{itemize}
The quantity $\Lambda_0\equiv\alpha_0/(\beta_0 \delta_0)=\gamma^{-1}\cdot\chi_{0}/(D_{c}t_{c,0})$ therefore characterizes the interplay between chemotatic and growth-driven spreading of bacterial populations. In particular, $\left[\chi_{0}/(D_{c}t_{c,0})\right]^{-1}$ is a characteristic time scale needed to spread \textit{via} chemotaxis over the nutrient diffusion length $\sqrt{D_{c}t_{c,0}}$, while $\gamma^{-1}$ is the time scale over which cells grow. Previous studies in bulk liquid focused solely on chemotactic spreading, which is characterized by the limit $\Lambda_0\gg1$ \cite{keller1971traveling,keller1975necessary,fu2018spatial,saragosti2011directional,seyrich2019traveling}. Other studies of non-chemotactic cells focused solely on growth-driven spreading, characterized by the opposite limit $\Lambda_0=0$ \cite{farrell2013mechanically,ben1998cooperative,eden1961two,murray2001mathematical,fujikawa1989fractal}. However, experiments performed in semi-solid agar \cite{croze2011migration} as well as in defined packings of particles \cite{bhattacharjee2021chemotactic} indicate that confinement in such crowded media introduces new cell-cell and cell-medium interactions that are not incorporated in the classic Keller-Segel model. Hence, in this paper, we describe a first step toward incorporating these complexities, which not only tune $\Lambda_0$ over a broad range, but also fundamentally alter spreading dynamics---as described hereafter.

\subsection{Keller-Segel model incorporating confinement}\label{Methods:KSModelWConfinement}
\noindent As a model system, we consider bacterial populations confined in media with close-packed, rigid, and immovable obstacles surrounding a free space that is sufficiently large for cells to move through. This form of confinement alters bacterial spreading dynamics in three ways: 
\begin{enumerate}[label=(\roman*)]
\item Collisions with the surroundings impede cellular migration \cite{cisneros2006reversal,drescher2011fluid,tapasoftmatter,051bhattacharjeenatcomm}, reducing the transport parameters $D_{b0}$ and $\chi_{0}$, as quantified in recent experiments in 3D porous media \cite{051bhattacharjeenatcomm,bhattacharjee2021chemotactic} as well as in semi-solid agar \cite{croze2011migration};
\item The presence of surrounding obstacles reduces the free space available to cells to move through, increasing cellular crowding and promoting cell-cell collisions that further truncate the motility parameters, observed experimentally using \textit{in situ} microscopy \cite{bhattacharjee2021chemotactic};
\item When the number density of cells is sufficiently high, this reduction in free space causes the cells to be jammed; hence, they are able to spread only through proliferation, which pushes cells outward, as quantified in experiments using single cell visualization \cite{dell2018growing,volfson2008biomechanical}.
\end{enumerate}
Notably, (ii)-(iii) are absent from the classic Keller-Segel model, which treats cells as non-contacting, and require modifications beyond simply changing the transport parameters $D_{b0}$ and $\chi_{0}$.

\begin{figure}[!h]
 \begin{center}
    \includegraphics[width=.6\textwidth]{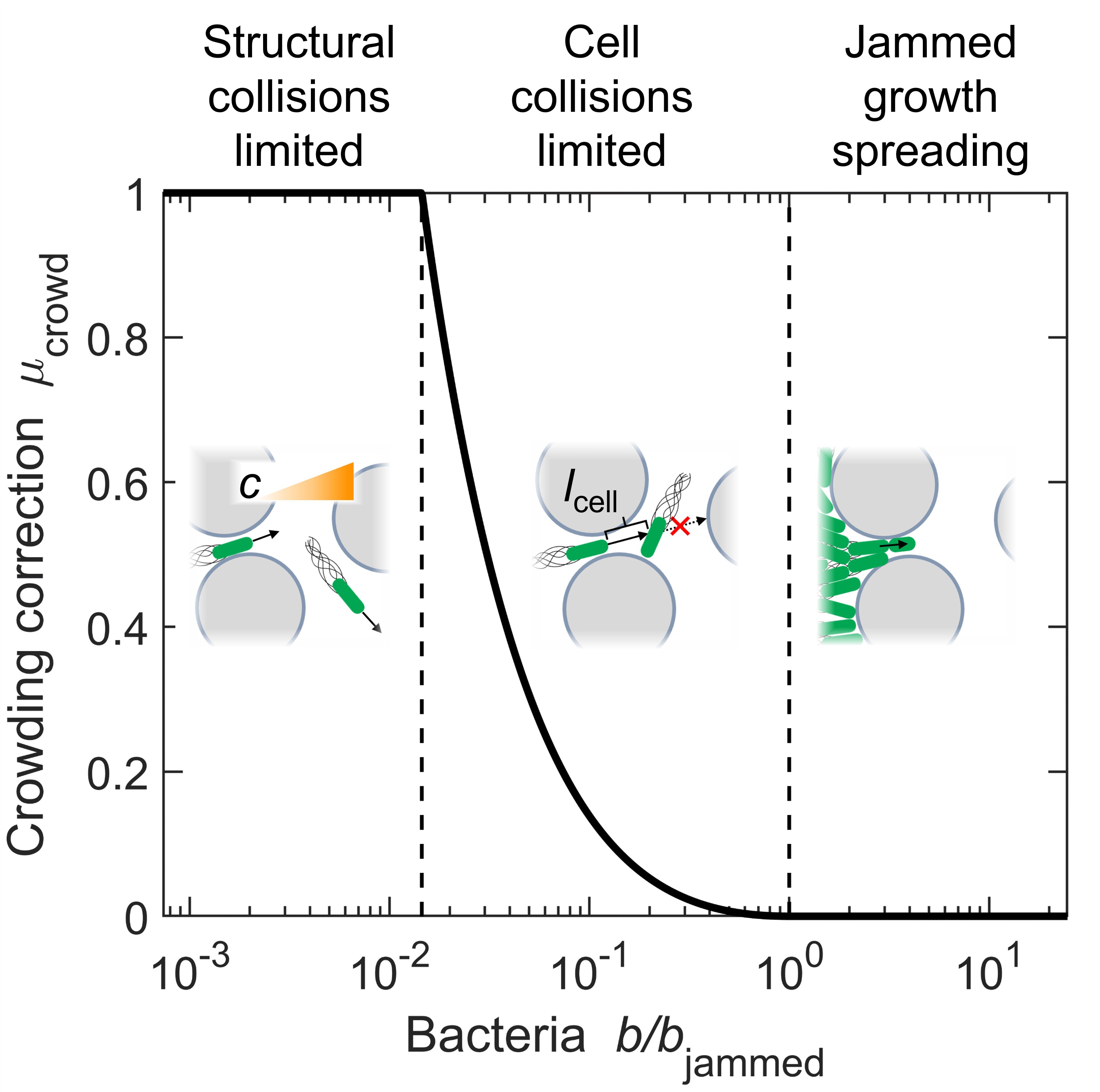}
  \end{center}
\caption{Summary of the cell density-dependent crowding correction $\mu_{\rm{crowd}}$, which we use in the model to incorporate the influence of confinement on cell-cell collisions. In particular, the cellular transport parameters are multiplied by $\mu_{\rm{crowd}}$. (Left) At low densities, migration of cells (green) is impeded only by collisions with surrounding solid obstacles (grey), not with neighboring cells, so $\mu_{\rm{crowd}}=1$. This impeded migration is quantified by the transport parameters $D_{b0}$ and $\chi_{0}$. Orange triangle shows a nutrient gradient along which cells bias their motion \textit{via} chemotaxis. (Middle) When the local density of cells is so large that the mean separation between neighboring cells $\ell_{\rm{cell}}$ is less than the mean chord length, which characterizes the amount of free space between obstacles, cell-cell collisions further truncate the transport parameters. This effect is quantified by $\mu_{\rm{crowd}}<1$. (Right) At the maximal density $b=b_{\rm{jammed}}$, the cells are jammed and have no free space to move. Therefore, $\mu_{\rm{crowd}}=0$, and the population spreads solely through growth and division of cells. Note that our definition of the number density of bacteria $b$ is as the number of cells per unit total volume of space, which includes the volume of surrounding obstacles.}
\label{fig:collisions_schematic}
\end{figure}

(i) \textit{Impeded migration of isolated cells}. Bacterial migration is typically modeled as a random walk with directed steps of characteristic length $\ell$ and characteristic duration $\tau$ that are punctuated by reorientation events \cite{bergecoli}. Consequently, both transport parameters $D_{b0}$, which describes the unbiased component of the random walk, and $\chi_0$, which describes the biased component, are set by $\sim \ell^{2}/\tau$. In bulk liquid, the directed steps are known as \textit{runs}, which extend along straight-line paths $\ell\sim40~\mu$m long, punctuated by rapid \textit{tumbles}. In tight confinement, however, a cell collides with an obstacle and becomes transiently trapped well before it completes such a run. Therefore, as established in recent experiments \cite{051bhattacharjeenatcomm,tapasoftmatter}, runs are truncated by collisions with surrounding obstacles, and the directed steps of the random walk are instead set by the geometry of the available free space; thus, for isolated cells, $\ell\sim\ell_{\rm{c}}$, the mean length of straight line \textit{chords} \cite{torquato} that fit in the free space \cite{tapasoftmatter}. Moreover, because the trapping process induced by collisions with obstacles occurs over a duration $\tau_{\rm{t}}$ that is longer than that of the truncated runs, $\tau\approx\tau_{\rm{t}}$. As a result, for cells confined in tight media, both transport parameters $D_{b0}$ and $\chi_0$ are instead $\propto \ell_{\rm{c}}^{2}/\tau_{\rm{t}}$---and because increasing confinement both decreases $\ell_{\rm{c}}$ and increases $\tau_{\rm{t}}$ \cite{tapasoftmatter}, it concomitantly decreases both $D_{b0}$ and $\chi_0$, as confirmed experimentally \cite{bhattacharjee2021chemotactic}.

(ii) \textit{Crowding-induced collisions between cells}. Confinement also reduces the free space available to cells. Our definition of the number density of bacteria $b$ quantifies the number of cells per unit total volume of space, which includes the volume of surrounding obstacles; hence, the local density of cells is given by $b/\phi$, where $\phi<1$ is the volume fraction of free space that is reduced by the presence of obstacles. This increase in the local density of cells increases the propensity of neighboring cells to collide as they move, further truncating $\ell$. Single-cell imaging in a porous medium confirms this expectation \cite{bhattacharjee2021chemotactic}: when the available free space is so tight that multiple cells cannot fit side-by-side, cells are necessarily restricted to end-on collisions between each other as they move, also inducing reorientations akin to those induced by collisions with surrounding obstacles. Therefore, as a first step toward incorporating this behavior into the Keller-Segel model, we adopt a mean-field treatment of cell-cell interactions in which cells truncate each other's directed steps in a density dependent manner, inducing transient trapping events again of duration $\tau_{\rm{t}}$ akin to collisions with obstacles. 

In particular, wherever the local density $b/\phi$ is larger than a threshold value $b^{*}/\phi$ such that the mean separation between the surfaces of neighboring cells, $\ell_{\text{cell}}$, decreases below the mean chord length $\ell_{\text{c}}$, we expect that cell-cell collisions truncate $\ell$ from $\ell_{\text{c}}$ to $\ell_{\text{cell}}$ (schematized in the middle inset of Fig. \ref{fig:collisions_schematic}). Because the diffusion and chemotactic coefficients both vary as $\propto \ell^{2}$, we therefore multiply both density-independent parameters $D_{b0}$ and $\chi_0$ that characterize isolated cells by the density-dependent correction factor $\mu_{\rm{crowd}}(b)=\left(\ell_{\text{cell}}/\ell_{\text{c}}\right)^{2}$, where the cell separation is approximated as the mean value $\ell_{\text{cell}}\equiv\left({3\phi}/{4 \pi b}\right)^{1/3}-d$; here, $d \approx 1~\mu$m is the characteristic size of a cell, and therefore $b^{*}\equiv{3\phi}/[{4 \pi  (\ell_{\text{c}}+d)^3}]$. As $b$ increases further, it eventually reaches the jamming density $b_{\rm jammed}\equiv{3\phi}/({4 \pi d^3})$ at which cells cannot move at all, and $\ell_{\text{cell}}=0$; in this case, both transport parameters are zero, and the bacterial population can only spread \textit{via} growth. Therefore, in Eq. \ref{eqnKellerSegelb}, $D_{b0}$ and $\chi_{0}$ are replaced by the corrected values
\begin{equation} \label{firstequation}
D_b(b)=D_{b0} \times \mu_{\rm{crowd}}(b)
\end{equation}
\begin{equation}\label{secondequation}
\chi(b)=\chi_{0} \times \mu_{\rm{crowd}}(b)
\end{equation}
where the crowding correction factor $\mu_{\rm{crowd}}(b)$ is piecewise defined as
\begin{equation} \label{crowdingEq}
\mu_{\rm{crowd}}(b)= \begin{cases} 
1 & \text{when} ~b \leq b^*\\ 
\left[\frac{\left(\frac{3\phi}{4\pi b}\right)^{1/3}-d}{\ell_{\rm{c}}}\right]^{2} & \text{when} ~b^* < b < b_{\rm{jammed}} \\ 
0 & \text{when} ~b \geq b_{\rm{jammed}} \\
\end{cases}
\end{equation}
as shown in Fig. \ref{fig:collisions_schematic}; the limits $b^*$ and $b_{\rm{jammed}}$ are indicated by the left and right vertical dashed lines, respectively. We term cases with low cell density ($b<b^*$) the \textit{obstacle collisions limited} regime described in (i) above; cases with intermediate cell density ($b^*\leq b< b_{\rm{jammed}}$) the \textit{cell collisions limited} regime; and cases with the highest possible density of cells ($b=b_{\rm{jammed}}$) the \textit{jammed growth spreading} regime described in (iii) below. Because we take the cells and surrounding obstacles to be incompressible, $b$ cannot exceed $b_{\rm{jammed}}$.

\begin{table*}[t]
\begin{adjustwidth}{-2.25in}{0in}
\centering
\begin{tabular}{|c|>{\centering\arraybackslash}m{0.12\linewidth}| >{\centering\arraybackslash}m{0.4\linewidth} |>{\centering\arraybackslash}m{0.14\linewidth}|}
\hline
\textbf{Parameter}  & \textbf{Weak confinement}         & \textbf{Intermediate confinement}         & \textbf{Strong confinement}        \\ \hline \rule{0pt}{10pt}
$\phi$         & 0.36         & 0.17           & 0.04           \\ \hline \rule{0pt}{10pt}

$\bar{l}_{\rm{c}}$ (\textmu{m})    & 4.6          & 3.1           & 2.4         \\ \hline \rule{0pt}{10pt}  

$D_{{b0}}$ $\big($\textmu{m}$^2$s$^{-1}\big)$   & 2.3  & 0.93  & 0.42          \\ \hline \rule{0pt}{10pt} 

$\chi_{\rm{0}}$ $\big($\textmu{m}$^2$s$^{-1}\big)$ & {3700} & {94}   & {16}         \\ \hline \rule{0pt}{10pt}

$\bar{b}/b_{\rm{max}}$    & {0.10}  & {0.027}      &{0.026}\\ \hline \rule{0pt}{10pt}

$\delta_0$    & {340}  & {860}      &{1900} \\ \hline \rule{0pt}{10pt}
$\delta(\bar{b})$    & {2.8$\times 10^{4}$}  & {1.2$\times 10^{4}$}      &{4.3$\times 10^{5}$} \\ \hline \rule{0pt}{10pt}

$\alpha=\alpha_0$    & {1600}          & {100}          & {38}              \\ \hline \rule{0pt}{10pt}

$\beta_0$     & {0.0063} & {0.0063 $(b_{0}=b_{\rm{max}})$}, {630 $(b_{0}=10^{-5}b_{\rm{max}})$} & {0.0063}\\ \hline \rule{0pt}{10pt} 

$\beta(\bar{b})$     & {0.06} & {0.23}  & {0.25}\\ \hline \rule{0pt}{10pt} 

$\Lambda_0$ & {750}   &  {18 $(b_{0}=b_{\rm{max}})$}, {1.8$\times 10^{-4}$ $(b_{0}=10^{-5}b_{\rm{max}})$} &{1.3}\\ \hline \rule{0pt}{10pt}

$\Lambda(\bar{b})$ & {0.95}   & {0.037}&{3.6 $\times 10^{-4}$}\\ \hline

\end{tabular}
\caption{Parameters used to describe bacteria in weak, intermediate, and strong confinement, as defined in the text. All parameters are defined in the text and their values are obtained from experiments as detailed in \S\ref{implementation} and the SI, with the exception of $\bar{b}$, which is determined directly from the simulation.}
\label{table_poreSizeDependent}
\end{adjustwidth}
\end{table*}
(iii) \textit{Jammed growth spreading.} When they are jammed, cells form a contact network that holds them in place and prevents motion by active propulsion. However, these cells can continue to proliferate if supplied with nutrient; based on the experiments in \cite{bhattacharjee2021chemotactic}, we assume that the maximal growth rate $\gamma$ is not affected by confinement. Thus, in this case, their high body stiffness enables growing cells to push outward on their neighbors; the bacterial population can then be treated as an incompressible ``fluid" in which the added stress due to cellular growth relaxes rapidly \textit{via} spreading, as is conventionally done in models of growing populations \cite{farrell2013mechanically,klapper2002finger,head2013linear} and supported by experiments \cite{dell2018growing,volfson2008biomechanical}. Because we treat the obstacles comprising the medium as being rigid and immovable, and the interstitial free space large enough for cells to move through without being deformed, this process leads to jammed growth spreading. To incorporate this behavior into the Keller-Segel model, at each time step $\delta_0 t$, we first identify the smallest $x_{i}$ at which $b(x_{i},t+\delta t)$ exceeds $b_{\rm jammed}$; we then set $b(x_{i},t+\delta t)=b_{\rm jammed}$ and instead relocate the newly-formed cells $\delta b(x_{i})\equiv b(x_{i},t+\delta t)-b_{\rm jammed}$ to the nearest location $x_{j}>x_{i}$ at which $b(x_{j},t)<b_{\rm jammed}$. We then repeat this process for all successive positions $x>x_{i}$ such that at time $t+\delta t$, the upper limit on cell density $b_{\rm jammed}$ is globally satisfied.

\subsection{Implementation of numerical simulations} \label{implementation}
\noindent To explore the influence of confinement, we perform numerical simulations of Eqs. \ref{eqnKellerSegelc}-\ref{eqnKellerSegelb}, modified as described in \S \ref{Methods:KSModelWConfinement}. Specifically, we implement a forward Euler method to solve these equations over time, discretizing the spatial coordinate $x$ using a forward difference form for first derivatives and a central difference form for second derivatives. The spatial resolution is 10 \textmu{m} and the time steps are 0.01 s; as shown in Fig. S1, these choices are sufficiently fine so that our results are not sensitive to
the choice of resolution. 

To connect our results to an experimental system, we use input parameters and initial conditions that mimic the experiments described in \cite{bhattacharjee2021chemotactic}, which explored the chemotactic migration of \textit{E. coli} populations in 3D porous media composed of densely-packed hydrogel particles. We use a Cartesian rectilinear coordinate system extending to a maximum distance of $1.75\times10^{4}$ \textmu{m}, matching the length of the experimental system. Because our system is one-dimensional, vectors (e.g. fluxes) oriented in the $+$ or $-x$ directions are represented by positive or negative quantities, respectively, with the vector notation suppressed. Both boundaries have no flux conditions. In these experiments, \textit{L}-serine was considered to act as the primary nutrient and chemoattractant for cells. Because the hydrogel particles are polymer networks swollen in liquid, they are permeable to the nutrient, similar to many other naturally-occurring media such as biological gels and microporous clays/soils. Therefore, we take the nutrient diffusivity $D_{{c}}$ to be equal to its value in bulk liquid, $800$ \textmu{m}$^2$ s$^{-1}$ \cite{ma2005studiesSerineDiffusion}, and the nutrient is initially saturated at $c_\infty=10~$mM throughout the simulation domain. For all the simulations, we use direct measurements of individual cells \cite{bhattacharjee2021chemotactic,cremer2019chemotaxis,croze2011migration} to choose fixed values of the cellular parameters $c_{-}$, $c_{+}$, and $\gamma$ given by $1$ \textmu{M}, $30$ \textmu{M}, and 0.69 h$^{-1}$, respectively; furthermore, as detailed in SI \S1, we use the data from experiments on spreading populations \cite{bhattacharjee2021chemotactic} to directly determine $c_{\rm{char}}$ and $\kappa$, given by 10 \textmu{M} and $1.3\times10^{-12}$ mM (cells/mL)$^{-1}$ s$^{-1}$, respectively.


Each experiment used a long 3D-printed cylinder of close-packed cells not containing hydrogel particles ($\phi=1$) as the initial inoculum, embedded within and surrounded by the 3D porous medium. The cells then continued to migrate radially outward through the pore space. Thus, as the initial condition in all the simulations, we consider a Gaussian profile of $b(x,t=0)$ centered at $x=0$ with a full width at half maximum of 100 \textmu{m} and a peak number density of $b_{0}=b_{\rm{max}}\equiv{3}/({4 \pi d^3})=2.4\times10^{11}$ cells/mL, where $b_{\rm{max}}$ is defined as the number density of close-packed cells and is therefore the maximum possible value of $b_{0}$ --- with the exception of the lower-density simulations presented in Figs. \ref{fig:profiles_dilute}-\ref{fig:transient_dilute}, which employ a lower value of $b_0$. For simplicity, wherever $b(x,t=0)>b_{\rm{jammed}}$, we still apply the jammed growth spreading rule described in \S  \ref{Methods:KSModelWConfinement}(iii), but with $b_{\rm{jammed}}$ replaced by $b(x,t=0)$. 

The experiments tuned cellular confinement by using porous media with varying porosities $\phi$ and mean chord lengths $\ell_{\rm{c}}$ \cite{tapasoftmatter}, resulting in varying values of the transport parameters $D_{b0}$ and $\chi_0$ \cite{bhattacharjee2021chemotactic,051bhattacharjeenatcomm}. In particular, as determined from the experiments, $D_{b0}$ and $\chi_0$ both decrease with increasing confinement as cellular mobility is increasingly hindered. Hence, in our simulations, we tune confinement by varying these parameters, using the values of $D_{b0}$ obtained from single-cell imaging \cite{bhattacharjee2021chemotactic} and extracting $\chi_0$ from experimental measurements of population spreading, as detailed in SI \S II. The confinement-dependent parameters are summarized in Table \ref{table_poreSizeDependent}.

The corresponding dimensionless parameters characterizing the Keller-Segel model (\S\ref{Methods:DimlessParam}) are also summarized in Table I:
\begin{itemize}

\item The diffusion parameter $\delta_0\equiv D_c/D_{b0}$ increases with confinement as cellular mobility is increasingly hindered. For all conditions tested here, however, $\delta_0$ is always much greater than one, reflecting fast diffusion of nutrient; thus, we expect that nutrient levels vary over large spatial extents, as confirmed in the simulations that follow. 

\item For all conditions tested here, the directedness parameter $\alpha_0\equiv\chi_0/D_{b0}$ is always much greater than one, indicating that motile cells strongly direct their motion in response to the nutrient gradient established through consumption. Intriguingly, the $\alpha_0$ determined from the experimental parameters decreases with increasing confinement, indicating that confinement more strongly hinders directed versus undirected motion --- consistent with previous reports that confinement fundamentally alters the mechanism by which cells perform chemotaxis \cite{bhattacharjee2021chemotactic}. 
Further investigating the determinants of $\alpha_0$ in confinement will be a useful direction for future experiments. 

\item Because the maximal growth rate is not affected by confinement \cite{bhattacharjee2021chemotactic}, the yield parameter $\beta_0\equiv\gamma/(b_{0}\kappa/c_{\infty})$ is independent of confinement for all of our simulations. For all simulations employing $b_{0}=b_{\rm{max}}$, $\beta_0$ is much less than one, reflecting the fact that nutrient consumption by a maximally dense population is faster than cellular proliferation; conversely, for the lower-density simulations presented in Figs. \ref{fig:profiles_dilute}-\ref{fig:transient_dilute}, $\beta_0$ is much greater than one, indicating the dominant role of proliferation in this case.

\end{itemize}
Therefore, for our simulations testing the influence of confinement on bacterial spreading, the parameter $\Lambda_0\equiv\alpha_0/(\beta_0 \delta_0)$ varies over a broad range, decreasing over nearly three orders of magnitude as confinement increases. We note that because the different parameters $\delta_0,\alpha_0,\beta_0$ do not incorporate the influence of density-dependent cellular crowding, we do not expect this transition to occur precisely at $\Lambda_{0}\approx1$. We therefore define a new version of this parameter, $\Lambda\equiv\alpha/(\beta \delta)$, where now $\delta\equiv D_{c}/D_{b}(\bar{b})$, $\alpha\equiv\chi(\bar{b})/D_{b}(\bar{b})=\alpha_0$, and $\beta\equiv\gamma/(\bar{b}\kappa/c_{\infty})$ (Table I); $\bar{b}$ is defined as the long-time mean cell density within each propagating pulse, and is directly calculated from each simulation as described further below. Thus, the newly-defined $\Lambda$ explicitly incorporates density-dependent crowding. As summarized in Table I, our simulations explore the transition from weak confinement ($\Lambda=0.95$) to strong confinement ($\Lambda=3.6\times10^{-4}$); consistent with our expectation, this range reflects a transition from chemotactic to growth-driven spreading, as demonstrated directly by the simulations presented below.

\begin{figure*}[!ht]
\includegraphics[width=\textwidth]{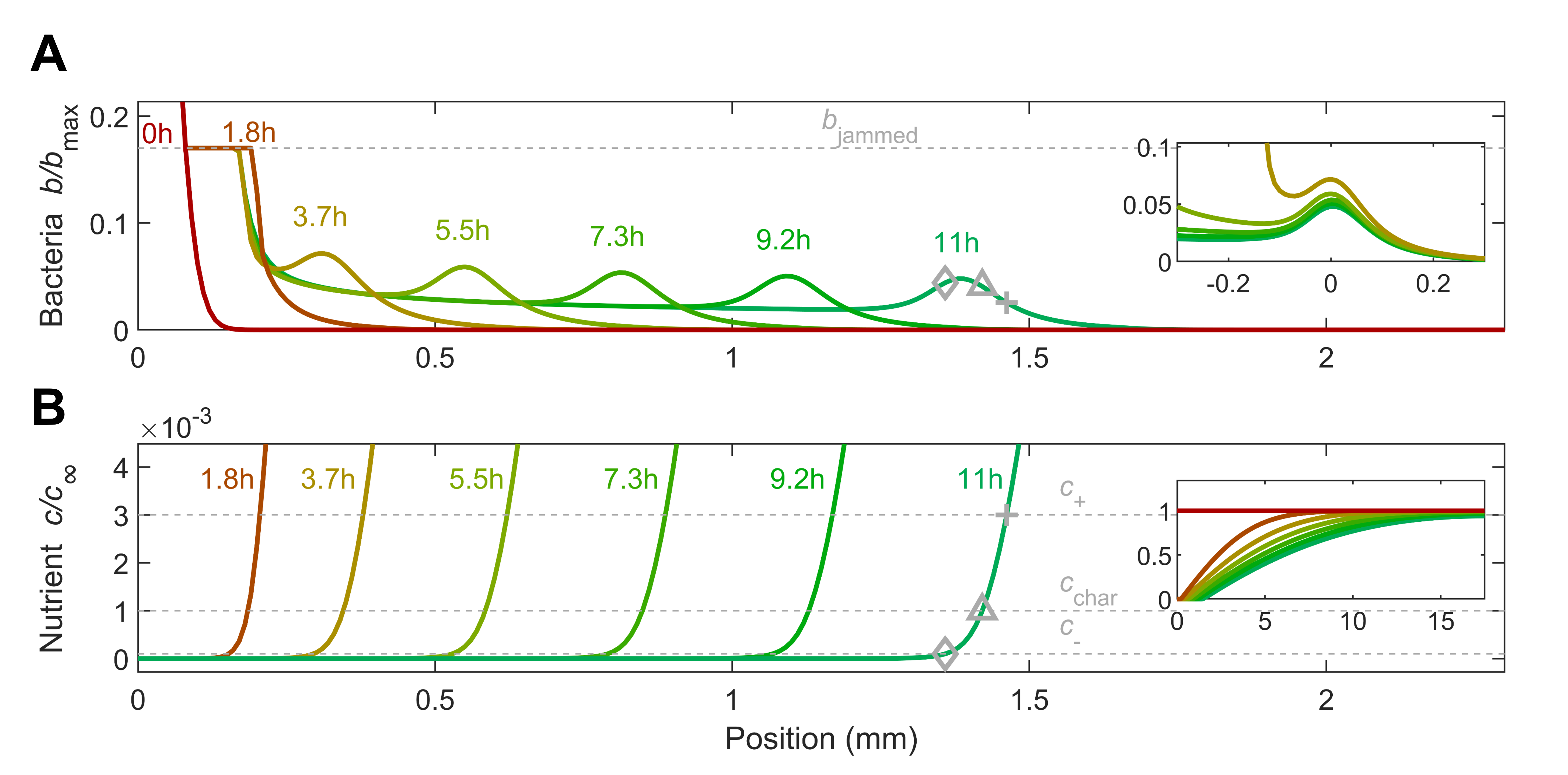}
\caption{Results from numerical simulations of population spreading in intermediate confinement. (A) shows the dynamics of the cells while (B) shows the corresponding dynamics of the nutrient, quantified by the normalized density $b/b_{\rm{max}}$ and concentration $c/c_{\infty}$, respectively. Different colors indicate different times as listed. The dense inoculum initially centered about the origin spreads outward, first as a jammed front (jamming density shown by the dashed grey line in A), then detaching as a coherent lower-density pulse that propagates continually \textit{via} chemotaxis. At long times, this pulse approaches an unchanging shape and speed, as shown by the collapse of the profiles in the upper inset (showing the same data, but shifted horizontally to center the peaks). The cellular dynamics arise in response to consumption of the nutrient, which is initially saturated everywhere, but is rapidly depleted and forms a gradient that is propagated with the pulse (inset shows the same data but with both axes zoomed out). In B, the three dashed grey lines show the characteristic concentrations of sensing $c_+$ and $c_-$ and the characteristic Monod concentration $c_{\rm{char}}$; the corresponding positions are shown by the pluses, diamonds, and triangles, respectively, in A-B. An animated form of this Figure is shown in Video S1.}
\label{fig:profiles_dense}
\end{figure*}

\section{Results}\label{Results}

\subsection{Intermediate confinement}\label{Results:intermediate}
\noindent As a prototypical starting case, we first examine bacterial spreading from a dense-packed Gaussian-shaped inoculum under intermediate confinement ($\Lambda=0.037$), shown by the initial profile for $t=0$ in Fig. \ref{fig:profiles_dense}A. 
The cells rapidly deplete nutrient locally \textit{via} consumption (Video S1) over a time scale $\sim c_{\infty}/\left(\kappa b_{\rm{max}}\right)\approx30$ s, establishing a steep nutrient gradient at the leading edge of the population. This gradient extends over a large distance ahead of the population (Fig. \ref{fig:profiles_dense}B and inset) --- as expected from our calculation of the diffusion parameter $\delta_0\gg1$. Cells at this leading edge then continue to grow outward as a jammed front with $b=b_{\rm jammed}$, shown by the flat region at $t=1.8$ h in Fig. \ref{fig:profiles_dense}A. Eventually, a lower-density, coherent pulse of cells detaches from this jammed region ($t=3.7$ h), continues to propagate the nutrient gradient along with it, and thus continues to migrate outward ($t>3.7$ h), as shown by the outward-moving peak in Fig. \ref{fig:profiles_dense}A.

Indeed, this pulse spans the extent over which nutrient varies between the upper and lower bounds of sensing, $c_+$ and $c_-$ (pluses and diamonds shown for the $t=11$ h profiles, respectively) --- reflecting the central role of chemotaxis in driving its propagation. The forward face of the pulse is also exposed to sufficient nutrient for cells to proliferate (with $c\geq c_{\rm{char}}$, the characteristic Monod concentration, shown by the triangles on the $t=11$ h profiles) --- suggesting that cellular growth contributes to population spreading over long time scales, as well. The overall width of this pulse, $W\approx200$ \textmu{m}, is set by the length scale over which nutrient is depleted by consumption; at its rear, the nutrient concentration and nutrient gradient are both low, causing both growth and chemotaxis to be hindered. As a result, cells are shed at a near-constant density $b_{\rm{trailing}}\approx0.02b/b_{\rm{max}}$ (see $0.5$ mm $<x<1.2$ mm in Fig. \ref{fig:profiles_dense}A). This coherent pulse of cells continues to move without an appreciable change in shape, as shown by the collapse in the inset to Fig. \ref{fig:profiles_dense}A, at a speed $v_{\rm{pulse}}\approx0.15$ mm/h. The nutrient profile concomitantly propagates with the pulse, as shown in Fig. \ref{fig:profiles_dense}B. Notably, similar spreading behavior was observed in experiments \cite{bhattacharjee2021chemotactic}.

\subsubsection{Initial dynamics}\label{Results:TransientDynamics}
\begin{figure*}
\includegraphics[width=\textwidth]{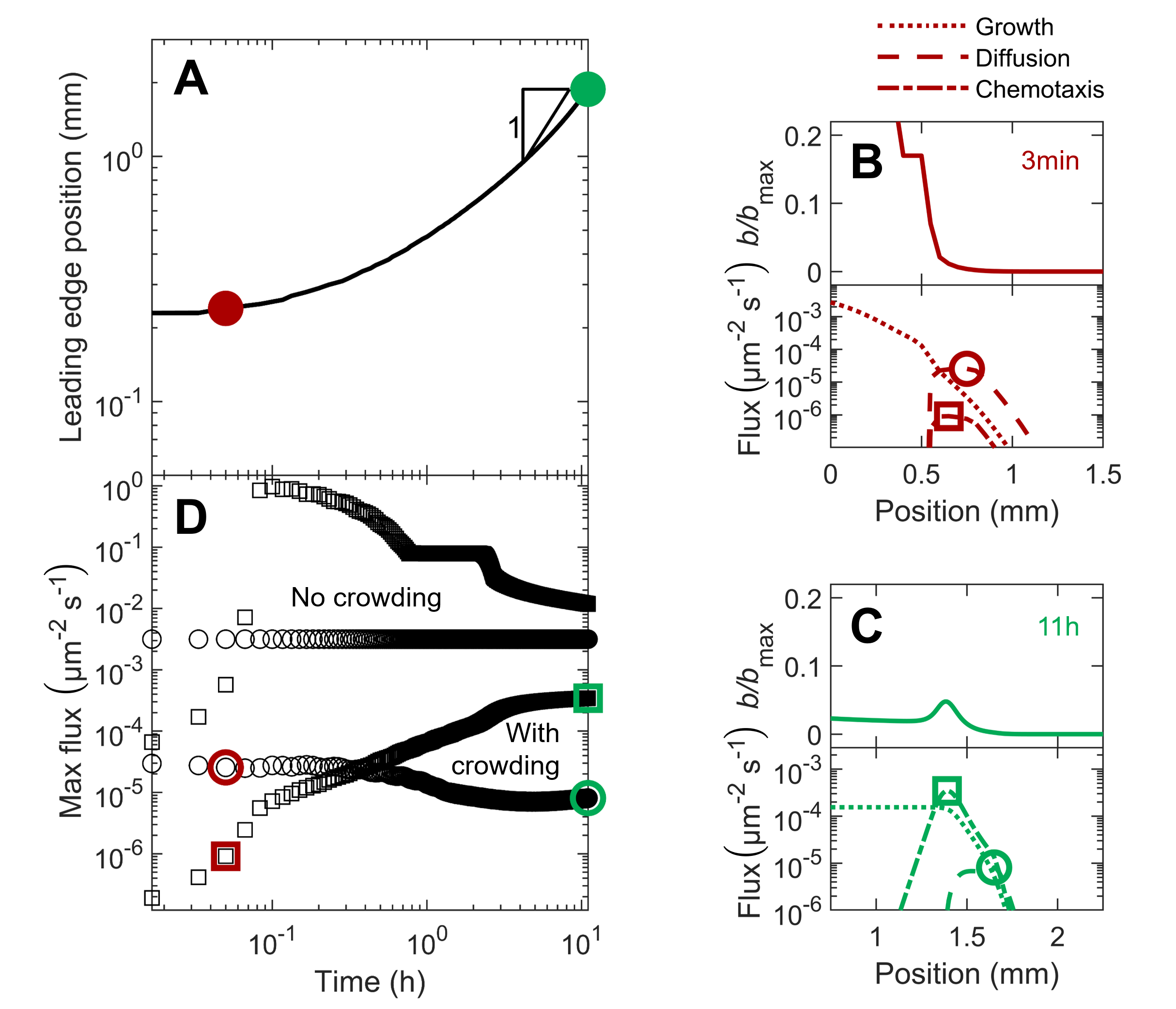}
\caption{Population dynamics, morphology, and fluxes driving spreading for simulations of bacteria in intermediate confinement (Fig. \ref{fig:profiles_dense}). (A) Increase in the position of the leading edge of the population is initially hindered (red), but approaches constant-speed motion indicated by the triangle at long times (green). (B) At a short time corresponding to the red point in A, the population expands as a jammed front (top panel). Lower panel shows that cellular growth and diffusion are the primary contributors to the expansion of this front. (C) At a long time corresponding to the green point in A, the population spreads as a coherent pulse (top panel). Lower panel shows that chemotaxis is the primary contributor to pulse propagation. Positions of the maximal diffusive and chemotactic fluxes are indicated by the circles and squares, respectively, in B-C; note the slight upward kink in the diffusive flux in C indicated by the circle. (D) Variation of the maximal diffusive and chemotactic fluxes, indicated by the circles and squares in B-C, over time. The initial population dynamics are dominated by cellular diffusion (circles), while at longer times chemotaxis dominates (squares). To illustrate the role played by cellular collisions, we show the same data with and without the crowding correction $\mu_{\rm{crowd}}$ in the upper and lower datasets; crowding hinders population spreading, as shown by the vertical offset in the curves, but plays a less appreciable role at long times, as shown by the curves approaching each other. }

\label{fig:transient_dense}
\end{figure*}

\noindent To further characterize these spreading dynamics, we track the position $x_{l}$ of the leading edge of the population over time $t$, as shown in Fig. \ref{fig:transient_dense}A. Initially, population spreading is hindered ($x_{l}\sim t^\nu$ with $\nu\ll1$ e.g., red point), but as the coherent pulse forms and propagates, it eventually approaches constant speed spreading ($\nu\approx1$ e.g., green point). A similar transition from hindered to constant speed spreading was observed in experiments \cite{bhattacharjee2021chemotactic}, although the underlying reason has thus far remained unclear. Here, we use our model to clarify the origin of this transition.


In particular, we examine the two different contributions to the motility-driven flux of cells -- active diffusion and chemotaxis -- for the population at early and late times (Figs. \ref{fig:transient_dense}B-C, respectively); for simplicity, we do not consider the added influence of growth, which only plays an appreciable role for long times $t\gg\gamma^{-1}$, until the next subsection. The magnitude of the active diffusive flux $-D_b(b)\nabla b=-D_{b,0}\mu_{\rm{crowd}}(b) \nabla b$ as it varies across the population is shown by the dashed lines in the bottom panels of Figs. \ref{fig:transient_dense}B-C, while the magnitude of the chemotactic flux $b v_c= b\chi_0 \mu_{\rm{crowd}}(b) \nabla f(c)$ is shown by the dash-dotted lines instead. At early times, the gradient in cell density is steep, as set by the sharp initial profile of cells and the limited extent of subsequent population spreading (Fig. \ref{fig:transient_dense}B, top). As a result, spreading is primarily due to active diffusion, which dominates over chemotaxis, as shown in the lower panel of Fig. \ref{fig:transient_dense}B. By contrast, as cells spread outward, the gradient in cell density becomes less steep. As a result, at late times, spreading is primarily due to chemotaxis, which dominates over active diffusion, as shown in the lower panel of Fig. \ref{fig:transient_dense}C. This behavior is also reflected by the bottom set of circles and squares in Fig. \ref{fig:transient_dense}D, which represent the maximal diffusive and chemotactic fluxes across the population (exemplified by the circles and squares in Figs. \ref{fig:transient_dense}B-C) over time. Initially, the diffusive flux dominates over the chemotactic flux; however, as the population continues to spread and consume nutrient, the diffusive flux decreases and the chemotactic flux increases, with both eventually approaching constant values at long times. 

Another key factor that hinders the initial population spreading is cellular crowding. To assess the influence of crowding, we compare the maximal diffusive and chemotactic fluxes across the population, but with or without the crowding correction factor $\mu_{\rm{crowd}}(b)$ (corresponding to the ``with crowding" and ``no crowding" datasets, respectively, in Fig. \ref{fig:transient_dense}D). In both cases, the active diffusive flux dominates over the chemotactic flux initially, but chemotaxis eventually dominates as the population continues to spread and establish the nutrient gradient (e.g. top set of squares for $t\leq0.1$ h). The spreading of the population remains hindered, however; due to the high initial density of cells, crowding continues to limit the chemotactic flux of cells, only enabling a small fraction at the leading edge of the population to migrate outward --- as exemplified by the first two profiles in Fig. \ref{fig:profiles_dense}A, the sharp decrease in both diffusive and chemotactic fluxes for $x<0.5$ mm in Fig. \ref{fig:transient_dense}B, and the large difference between the two sets of squares in Fig. \ref{fig:transient_dense}D. Eventually, as this leading edge continues to migrate, crowding in the forward face of the population becomes sufficiently low, enabling the coherent pulse of cells to detach from the population --- as exemplified by the $t=3.7$ h profile in Fig. \ref{fig:profiles_dense}A and the ``kink" in the top set of squares at $t\approx3$ h in Fig. \ref{fig:transient_dense}D. Hindrance due to crowding continues to decrease over time, as shown by the diminishing difference between the two sets of squares in Fig. \ref{fig:transient_dense}D for $t>3$ h, and eventually approaches a constant value. 

Hence, population spreading is initially slow due to the time required for cellular consumption to establish a sufficiently strong nutrient gradient to drive chemotactic migration. Cellular crowding near the initial inoculum then continues to hinder spreading until enough of the forward face of the population has migrated outward --- enabling cells to detach as a coherent pulse that continues to move outward, eventually approaching a constant speed.

\subsubsection{Long-time behavior}\label{Results:SteadyStateBehavior}
\noindent Having established how the spreading population forms a moving pulse, we now seek to clarify the factors that continue to drive its propagation. As previously described (Fig. \ref{fig:transient_dense}), active diffusion plays a negligible role at these longer times. Instead, as noted previously when describing the $t=11$ h profiles in Fig. \ref{fig:profiles_dense}, we expect that chemotaxis and growth are the principal contributors to population spreading. In particular, the outward-moving pulse spans the extent over which nutrient varies between the upper and lower bounds of nutrient sensing --- reflecting the central role of chemotaxis in driving its propagation. The forward face of the pulse is also exposed to sufficient nutrient for cells to proliferate --- suggesting that cellular growth contributes to spreading, as well. Indeed, the time scale over which this pulse propagates over its width $\sim W/v_{\rm{pulse}}= 1.3$ h is comparable to the time scale of cellular proliferation, $\gamma^{-1}=1.4$ h, further indicating that growth may contribute to population spreading. However, the relative influence of chemotaxis versus growth in driving population spreading remains unclear. 

To address this gap in knowledge, we examine the long-time behavior of the pulse by considering a coordinate system that moves with the pulse, $\xi\equiv x-tv_{\rm{pulse}}+\xi_0$; $\xi_0$ is a constant shift factor chosen such that $\xi=0$ is located at the rear of the pulse, at which $b\approx b_{\rm{trailing}}$. Here, both the bacterial and nutrient gradients are negligible, eliminating diffusive and chemotactic fluxes of cells, as shown in Fig. \ref{fig:transient_dense}C. Within a time increment $dt$, the moving pulse leaves behind $N_{\rm{loss}}\approx b_{\rm{trailing}}v_{\rm{pulse}} A dt$ cells, where $A$ is the transverse cross sectional area. Simultaneously, growth generates $N_{\rm{grown}}\approx A dt \int_{\xi=0}^\infty b(\xi')\gamma g(\xi') d\xi'$ new cells within the pulse. Therefore, $N_{\rm{loss}}\approx N_{\rm{grown}}$ to preserve a nearly-unchanging pulse of cells (Fig. \ref{fig:profiles_dense}A, inset). 

More generally, at locations further ahead ($\xi\geq0$), $N_{\rm{motile}}\approx J_{m}Adt$ cells also travel with the pulse through their motility-driven flux $J_{m}=-D_b\nabla b+b v_c$; here, $b$, $\nabla b$, $D_{\rm{b}}$, and $v_{\rm{c}}$ are all $\xi$-dependent quantities. Thus, an unchanging profile of cells requires the more general flux balance $N_{\rm{loss}}-N_{\rm{motile}}\approx N_{\rm{grown}}$, where now $N_{\rm{loss}}\approx bv_{\rm{pulse}}Adt$ and $N_{\rm{grown}}\approx A dt \int_\xi^\infty b(\xi')\gamma g(\xi') d\xi'$; that is, 
\begin{equation}\label{eqnBflux}
   \underbrace{b v_{\rm pulse}}_{\text{Loss}}  \   \ + {\underbrace{D_{\rm{b}} \nabla b }_{\text{Diffusion}}} \ \ -  \underbrace{bv_{\rm{c}}}_{\text{Chemotaxis}}\approx \  \underbrace{\int_\xi^\infty b\gamma g d\xi' }_{\text{Growth}} 
\end{equation}
where all quantities except for the constants $v_{\rm pulse}$ and $\gamma$ are position-dependent. This equation quantifies the intuition that the cells that cannot keep up with the moving pulse through their motility must be replaced by growth so as to prevent a net loss of cells from the region ahead of $\xi$. Therefore, for a given position $\xi$, the right hand side of Eq. \ref{eqnBflux} represents the additional contribution to the overall spreading of the pulse due to cellular growth at $\xi\geq0$. We therefore term this quantity the \textit{growth flux} and compare it to the chemotactic flux $bv_{\rm{c}}$.

Both fluxes are shown for the final profile in Fig. \ref{fig:transient_dense}C; the growth flux is shown by the dotted line and the chemotactic flux is shown by the dash-dotted line, both plotted on a logarithmic scale. A version showing these fluxes on linear scales is shown in Fig. \ref{fig:combinedconfinement}E. For this case of intermediate confinement, both fluxes are appreciable, with the maximal chemotactic flux ($3.4\times 10^{-4}$ cells \textmu{m}$^{-2}\rm{s}^{-1}$) slightly larger than the maximal growth flux ($1.5\times 10^{-4}$ cells \textmu{m}$^{-2}\rm{s}^{-1}$), indicating that chemotaxis plays a greater role in driving population spreading. To further quantify this behavior, we evaluate Eq. \ref{eqnBflux} at two distinct positions: the rear of the pulse ($\xi=0$) and the peak of chemotactic flux, which we denote $\xi_{\rm{peak}}$ (indicated by the square in Fig. \ref{fig:transient_dense}C). At both locations, the gradient in cell density is approximately zero, eliminating diffusive flux and simplifying our analysis. The chemotactic flux is also approximately zero at the rear of the pulse ($x\approx1.2$ mm in Fig. \ref{fig:transient_dense}C). Moreover, at both locations, the growth flux is approximately the same --- reflecting the fact that only the forward face of the pulse is exposed to sufficient nutrient for cells to proliferate. Hence, equating both of these implementations of Eq. \ref{eqnBflux} yields an expression for the long-time pulse speed:
\begin{equation}\label{eqnVelocityRelationship}
v_{\rm pulse}\approx v_c(\xi_{\rm{peak}})+v_{\rm pulse}\frac{b_{\rm{trailing}}}{b(\xi_{\rm{peak}})}.
\end{equation}
Therefore, the ratio ${b_{\rm{trailing}}}/{b(\xi_{\rm{peak}})}=40\%$ approximates the fraction of the overall pulse speed attributable to growth, while the remaining $60\%$ is due to chemotaxis.

This analysis also provides a way to extend a previous scaling estimate \cite{cremer2019chemotaxis} of the long-time pulse speed $v_{\rm pulse}$, which did not incorporate the influence of confinement in regulating spreading. First, we note that the chemotactic velocity scales as $v_c(\xi_{\rm{peak}})\sim\chi\left(b(\xi_{\rm{peak}})\right)/W$, where $W$ is the pulse width. Next, we relate the mean number density of cells $\bar{b}\equiv W^{-1}\int_{0}^{\infty}{b d\xi'}$ to $v_{\rm pulse}$ through a flux balance of cells at long times, when the shape of the pulse is unchanging over time. In particular, as described earlier, the rate at which cells are left behind the pulse, $b_{\rm{trailing}}v_{\rm pulse}A$, is balanced by the rate at which growth generates new cells in the pulse, $A\int_{0}^{\infty}{b\gamma g d\xi'}=A\bar{b}W\gamma\bar{g}$, where we have defined the cell-weighted mean $\bar{g}\equiv \int_{0}^{\infty} g(c(\xi')) b d\xi' /\int_{0}^{\infty} b d\xi' =\int_{0}^{\infty} g(c(\xi')) b d\xi'/(\bar{b}W) $. This flux balance yields $W=b_{\rm{trailing}}v_{\rm pulse}/\left(\bar{b}\gamma\bar{g}\right)$, and therefore, $v_c(\xi_{\rm{peak}})\sim\chi\left(b(\xi_{\rm{peak}})\right)\bar{b}\gamma\bar{g}/\left(b_{\rm{trailing}}v_{\rm pulse}\right)$. Substituting this expression into Eq.~\ref{eqnVelocityRelationship},
\begin{equation}
 v_{\rm pulse}\approx \chi\left(b(\xi_{\rm{peak}})\right)\frac{\bar{b}\gamma \bar{g}(c)}{b_{\rm{trailing}}v_{\rm pulse}} + v_{\rm pulse}\frac{b_{\rm{trailing}}}{b(\xi_{\rm{peak}})},
\end{equation}
which can then be rearranged to yield our ultimate scaling estimate for $v_{\rm pulse}$:
\begin{equation}\label{result}
 v_{\rm pulse}^2\approx \chi\left(b(\xi_{\rm{peak}})\right) \gamma \bar{g}(c) \frac{\bar{b}b_{\rm{peak}}}{b_{\rm{trailing}}(b_{\rm{peak}}-b_{\rm{trailing}})}.
\end{equation}
This estimate thus extends a previous calculation \cite{cremer2019chemotaxis} by explicitly incorporating the influence of confinement. To evaluate the accuracy of this estimate, we use the long-time simulation data to directly determine all the parameters on the right hand side of Eq.~\ref{result} and thereby obtain $v_{\rm pulse}$. We find reasonable agreement between the predicted (\textit{via} Eq.~\ref{result}) and simulated speeds to within a factor of two: the predicted value is 0.08 mm/h, while the simulation yields 0.15 mm/h. Hence, Eq.~\ref{result} provides a straightforward way to approximately relate the long-time shape of a pulse to its propagation speed, even in confinement.

Finally, we note that the fluxes associated with chemotaxis and growth also determine the overall shape of the spreading population; for simplicity, we neglect the diffusive flux, given that it is at least one order of magnitude smaller than the chemotactic and growth fluxes (Fig. \ref{fig:transient_dense}C). In particular, as quantified in Eq. \ref{eqnBflux}, the cellular profile $b(\xi)$ is given by the sum of the chemotactic and growth fluxes, scaled by the constant $v_{\rm{pulse}}$. Our results confirm this expectation: as shown in Fig. \ref{fig:transient_dense}C, the location of the bacterial pulse nearly coincides with the peak in the chemotactic flux, while the steady increase in growth flux from the leading edge to the rear coincides with the additional asymmetry in the bacterial profile arising from the trail of cells shed from the moving pulse. Taken together, these results therefore demonstrate that the interplay between chemotaxis and growth determines both the long-time speed and shape of the spreading population.

\begin{figure*}[!ht]
\centering
\includegraphics[width=1\textwidth]{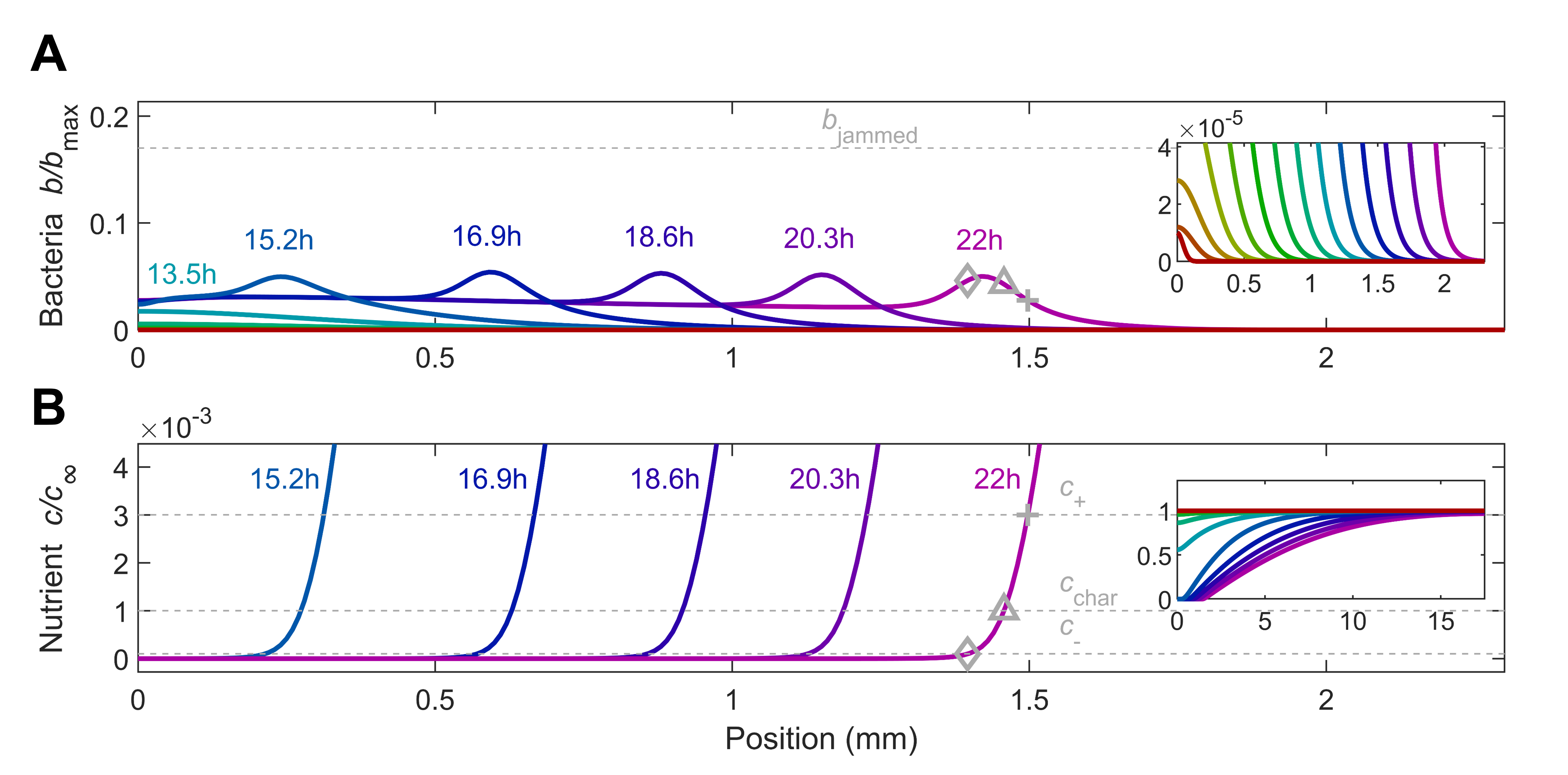}
\caption{Results from numerical simulations of population spreading in intermediate confinement starting from a more dilute inoculum. (A) shows the dynamics of the cells while (B) shows the corresponding dynamics of the nutrient, quantified by the normalized density $b/b_{\rm{max}}$ and concentration $c/c_{\infty}$, respectively. Different colors indicate different times as listed. The dilute inoculum (jamming density shown by the dashed grey line in A) initially centered about the origin first grows exponentially and spreads diffusively until nutrient is locally depleted (upper inset shows the same data, but zoomed in to the vertical axis); only then does a coherent pulse detach and propagate continually \textit{via} chemotaxis in response to the nutrient gradient (lower inset shows the same data but with both axes zoomed out). Even though the short-time behavior is different from the case of a more dense inoculum shown in Fig. \ref{fig:profiles_dense}, the long-time behavior of this pulse is identical. To facilitate comparison with Fig. \ref{fig:profiles_dense}, in B, the three dashed grey lines again show the characteristic concentrations of sensing $c_+$ and $c_-$ and the characteristic Monod concentration $c_{\rm{char}}$; the corresponding positions are shown by the pluses, diamonds, and triangles, respectively, in A-B. An animated form of this Figure is shown in Video S2.}
\label{fig:profiles_dilute}
\end{figure*}

\begin{figure*}
\centering
\includegraphics[width=\textwidth]{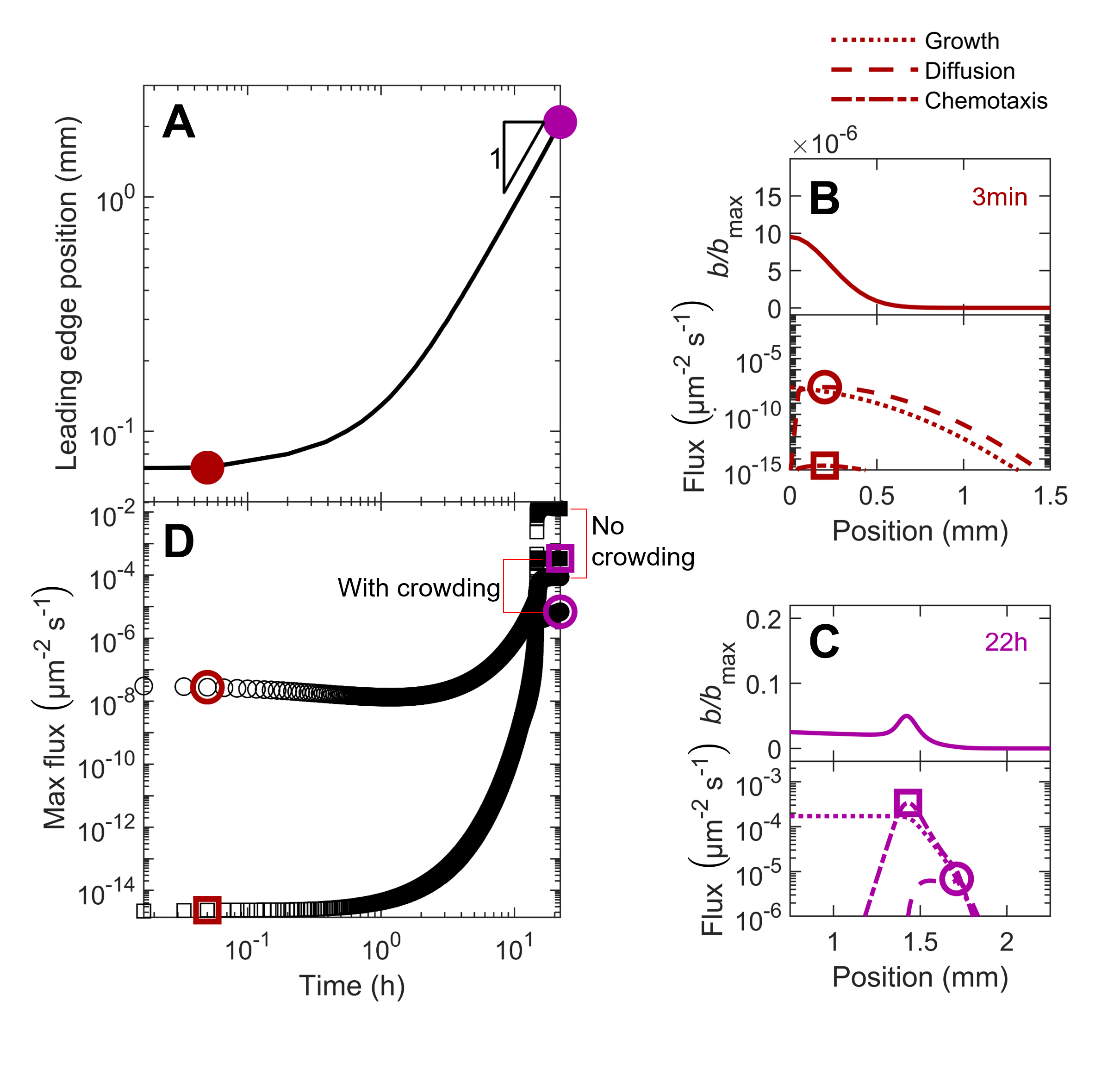}

\caption{Population dynamics, morphology, and fluxes driving spreading for simulations of bacteria in intermediate confinement (Fig. \ref{fig:profiles_dilute}). (A) Increase in the position of the leading edge of the population is initially hindered (red), but approaches constant-speed motion indicated by the triangle at long times (purple). (B) At a short time corresponding to the red point in A, the population grows exponentially (top panel) and spreads primarily through growth and diffusion (lower panel). (C) At a long time corresponding to the purple point in A, the population spreads as a coherent pulse (top panel). Lower panel shows that chemotaxis is the primary contributor to pulse propagation. Positions of the maximal diffusive and chemotactic fluxes are indicated by the circles and squares, respectively, in B-C; note the slight upward kink in the diffusive flux in C indicated by the circle. (D) Variation of the maximal diffusive and chemotactic fluxes, indicated by the circles and squares in B-C, over time. The initial population dynamics are dominated by cellular diffusion (circles), while at longer times chemotaxis dominates (squares). To illustrate the role played by cellular collisions, we show the same data with and without the crowding correction $\mu_{\rm{crowd}}$ in the datasets indicated by the red lines; the data are identical except at long times, when crowding slightly hinders population spreading. }

\label{fig:transient_dilute}

\end{figure*}

\subsubsection{Influence of initial cell density}\label{Results:InfluenceStartingDensity}
\noindent Our analysis thus far considered a dense initial inoculum, for which cellular crowding hinders the formation and detachment of a pulse; at much longer times, this less-crowded pulse no longer resembles the initial inoculum, but instead is shaped by the interplay of chemotaxis and growth (Fig. \ref{fig:transient_dense}). We therefore expect that for a lower-density inoculum, a similar pulse also emerges at long times, but with initial dynamics that are limited instead by the time required for cellular consumption to establish a sufficiently strong nutrient gradient. To test this expectation, we repeat the simulation shown in Fig. \ref{fig:profiles_dense}, but using an initial peak number density of cells that is $10^{5}$ times smaller ($b_{0}=10^{-5}b_{\rm{max}}$) --- shown in Video S2.

In the previously-considered case of a dense inoculum, the cells deplete nutrient rapidly \textit{via} consumption, and the population subsequently spreads from its leading edge as a growing jammed front (first two curves in Fig. \ref{fig:profiles_dense}). By contrast, with a more dilute inoculum, nutrient depletion takes much longer. Instead, the population continually grows and spreads as a whole (red to blue curves in Fig. \ref{fig:profiles_dilute}A inset), not just at its leading edge, without appreciably depleting nutrient. It eventually reaches a maximal density $b'=b_{0}e^{\gamma t'}$ for which the time scale of subsequent nutrient depletion $t_{\rm{dep}}\sim c_{\infty}/\left(\kappa b'\right)$ is comparable to the time scale of subsequent growth $t_{\rm{g}}\sim\gamma^{-1}$; equating these time scales yields $t'=\gamma^{-1}\ln{\left(\frac{c_\infty \gamma}{b_0 \kappa}\right)}$. Therefore, we expect that nutrient is fully depleted at the initial inoculum after $t'+t_{\rm{dep}}\sim\gamma^{-1}\left[\ln{\left(\frac{c_\infty \gamma}{b_0 \kappa}\right)}+1\right]\approx11$ h. The simulation results are consistent with this estimate, which neglects spatial variation in nutrient availability through the entire population and thus serves as a lower bound, showing that nutrient is fully depleted at the initial inoculum after $\sim14$ h (red to blue curves in Fig. \ref{fig:profiles_dilute}B inset). The nutrient gradient again extends over a large distance ahead of the population, as expected from our calculation of the diffusion parameter $\delta_0\gg1$.

Unlike the case of a dense inoculum, the population does not subsequently spread as a jammed front. Instead, once the nutrient gradient is sufficiently strong, a coherent pulse of cells again detaches \textit{without} the prior formation of a jammed front, continues to propagate the nutrient gradient with it, and continues to migrate outward ($t>15$ h in Fig. \ref{fig:profiles_dilute}). Consistent with our expectation, this pulse is noticeably similar to that which arises in the dense inoculum case: it has a nearly-identical shape and also moves without an appreciable change in shape, eventually reaching approximately the same constant speed, $v_{\rm{pulse}}\approx0.16$ mm/h (compare late-time profiles in Figs. \ref{fig:profiles_dense} and \ref{fig:profiles_dilute}).

To further characterize the population spreading dynamics, we again plot the leading edge position $x_{\ell}$ as a function of time $t$. As in the case of a dense inoculum, $x_{\ell}\sim t^{\nu}$ with $\nu\ll1$ at early times, transitioning to $\nu\approx1$ at later times (Fig. \ref{fig:transient_dilute}A); however, these seemingly similar dynamics reflect fundamentally different underlying processes at early times. With a more dilute inoculum, slower nutrient depletion causes the diffusive flux to initially dominate over chemotaxis (Fig. \ref{fig:transient_dilute}B) \textit{without} any influence of cellular crowding --- indicated by the overlap of the early-time points with/without the crowding correction in Fig. \ref{fig:transient_dilute}D. As cells continue to grow and consume nutrient, they eventually establish a sufficiently strong gradient and spread as a coherent pulse \textit{via} chemotaxis --- as indicated by the dominant role of the chemotactic flux at long times (Fig. \ref{fig:transient_dilute}C-D). At these later times, the different contributions to the bacterial flux are nearly identical to those that drive pulse propagation in the case of a dense inoculum (compare Figs. \ref{fig:transient_dilute}C and \ref{fig:transient_dense}C). Indeed, the fractions of the overall pulse speed attributable to chemotaxis and growth, as quantified by Eq. \ref{eqnVelocityRelationship}, are $\approx58\%$ and $42\%$, respectively --- nearly identical to the case of a dense inoculum. Hence, while the initial dynamics of population spreading are sensitive to the initial cell density -- consistent with experiments \cite{bhattacharjee2021chemotactic} -- the properties of the pulse that forms and continues to drive spreading at long times are not, instead being set solely by the interplay between chemotaxis and growth.

\subsection{Influence of confinement}\label{Results:InfluenceConfinement}

\begin{figure*}[!ht]
\centering
\includegraphics[width=1\linewidth]{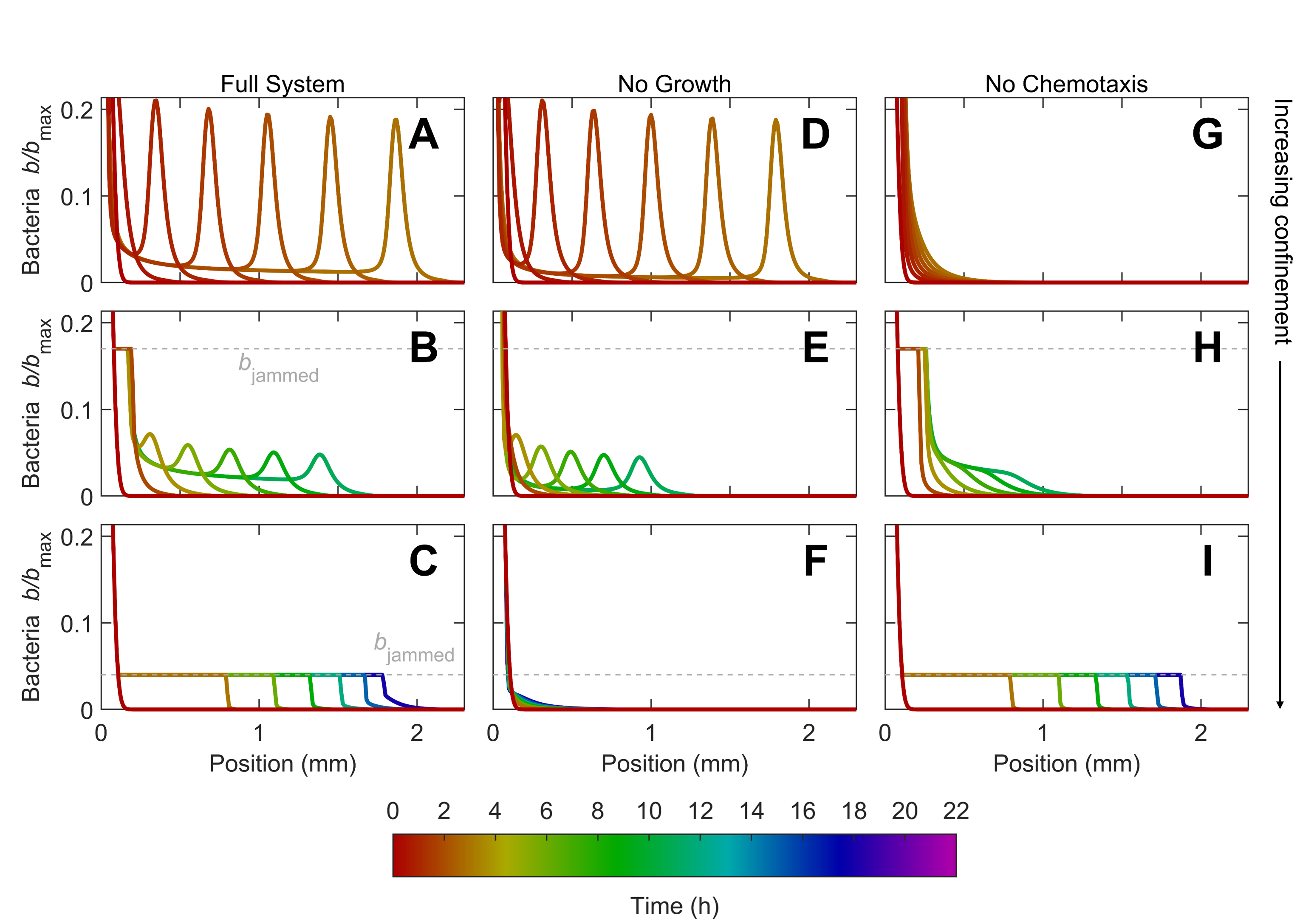}
\caption{Increasing confinement causes a transition from fast chemotactic pulse propagation to slower jammed growth expansion. Panels show results from numerical simulations of population spreading from the same dense inoculum initially centered about the origin in weak, intermediate, and strong confinement, shown by top, middle, and bottom rows respectively. First column shows the results of the full model, while second and third columns show the same simulations with growth or chemotaxis omitted, respectively. We only show the normalized cellular density $b/b_{\rm{max}}$ for clarity. Different colors indicate different times as listed in the color scale. In weak confinement, a coherent pulse rapidly detaches and continually propagates; this pulse is driven primarily by chemotaxis, and thus, omitting growth barely changes the dynamics while omitting chemotaxis abolishes the propagation altogether. Conversely, in strong confinement, the population spreads slowly as a jammed front, driven primarily by growth. In intermediate confinement, both growth and chemotaxis drive population spreading. The dashed grey line shows the jamming density, which varies depending on confinement (and is larger than the vertical scale in the top row). An animated form of this Figure, along with the nutrient profiles, is shown in Video S3.}
\label{fig:knockouts}

\end{figure*}

\begin{figure}
    \centering
    \includegraphics[width=1\textwidth]{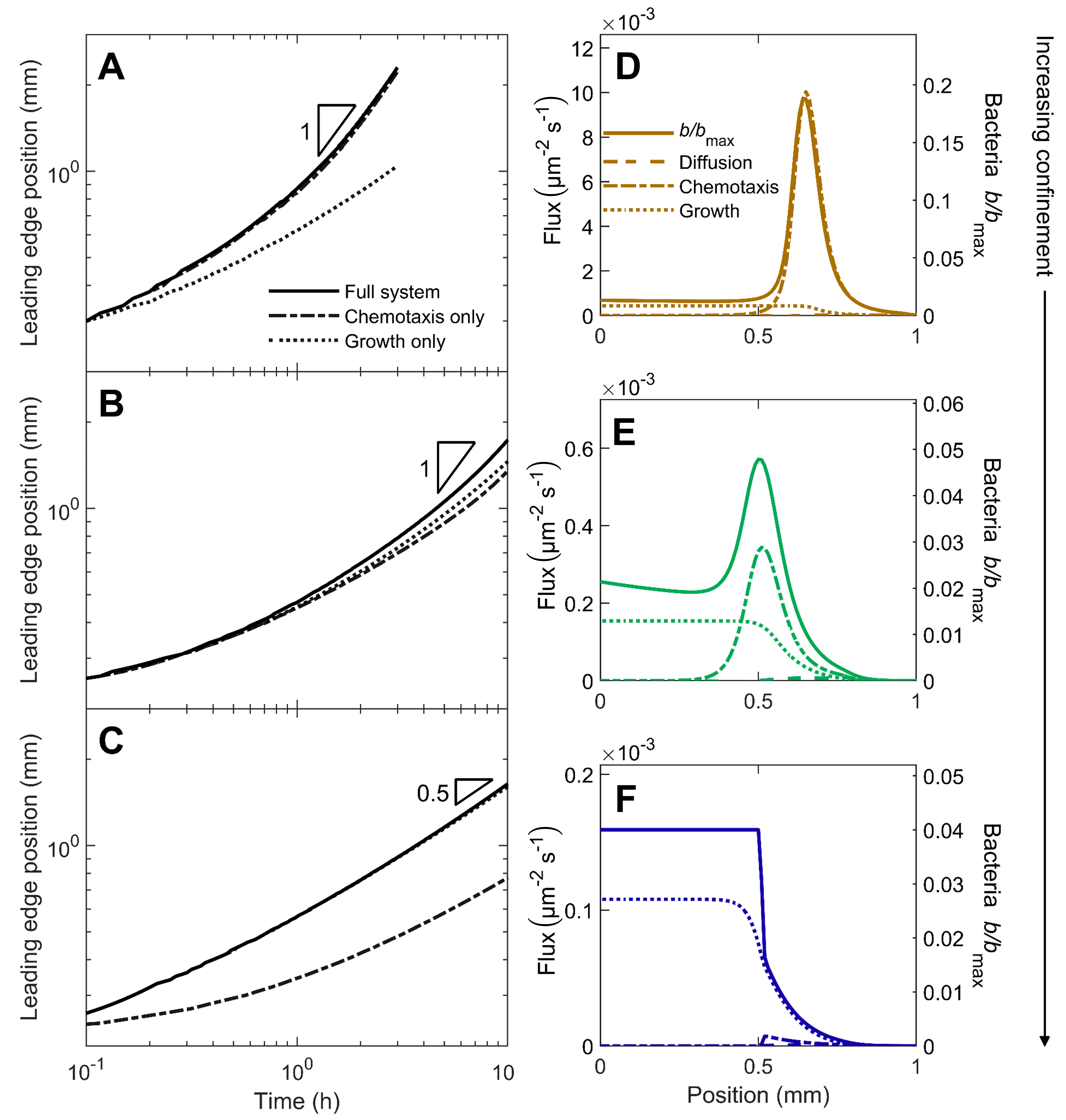}
\caption{Increasing confinement causes a transition from fast chemotactic pulse propagation to slower jammed growth expansion. Panels quantify population dynamics, morphology, and fluxes driving spreading for simulations of bacteria in weak, intermediate, and strong confinement (Fig. \ref{fig:knockouts}) as shown by the top, middle, and bottom rows, respectively. (A-B) Increase in the position of the leading edge of the population is initially hindered, but approaches constant-speed motion indicated by the triangle at long times. In strong confinement (C), however, the long-time behavior approaches diffusive-like scaling instead. (D-E) At long times, the population spreads as a coherent pulse (solid line) driven primarily by chemotaxis in weak confinement, and by both chemotaxis and growth in intermediate confinement. (F) In strong confinement, however, the population spreads slowly as a jammed front, driven primarily by growth. }
    \label{fig:combinedconfinement}
\end{figure}

\noindent For the case of intermediate confinement explored thus far, we have established that chemotaxis and growth both drive population spreading at long times. How does this behavior change with confinement? As quantified in Figs. \ref{fig:transient_dense}D and \ref{fig:transient_dilute}, confinement-induced crowding limits the chemotactic flux; therefore, we expect that with reduced or increased confinement, chemotaxis or growth plays a more dominant role in driving spreading, respectively. To test this expectation, we perform the same simulation with a dense inoculum as in Figs. \ref{fig:profiles_dense}-\ref{fig:transient_dense}, but with different values of the confinement-dependent parameters as summarized in Table \ref{table_poreSizeDependent}. In particular, our simulations explore $\Lambda=0.95$, $0.037$, and $3.6\times10^{-4}$, representing weak, intermediate, and strong confinement (top, middle, and bottom rows in Figs. \ref{fig:knockouts}--\ref{fig:combinedconfinement}), respectively --- also shown in Video S3. 

In all cases, the cells first rapidly deplete nutrient locally \textit{via} consumption, generating a nutrient gradient that again extends over a large distance and drives subsequent spreading at the leading edge of the population. However, consistent with our expectation, and with experimental observations \cite{bhattacharjee2021chemotactic}, the nature of this spreading is strongly confinement-dependent. 

\subsubsection{Weak confinement}\noindent In the case of weak confinement, cells detach and spread as a lower-density, coherent, propagating pulse \textit{without} first growing outward as a jammed front (Fig. \ref{fig:knockouts}A), unlike the case of intermediate confinement (Fig. \ref{fig:knockouts}B). This pulse is notably sharper and faster, with the long-time pulse speed and peak height $\approx5.4$ and $3.9$ times larger than in intermediate confinement (also compare Panels A--B and D--E in Fig. \ref{fig:combinedconfinement}) --- reflecting the dominant role of chemotaxis in driving spreading, as expected from the larger value of $\Lambda$. Quantification of the different fluxes driving spreading corroborates this expectation (Fig. \ref{fig:combinedconfinement}D); indeed, following our previous analysis summarized by Eqs. \ref{eqnBflux}-\ref{eqnVelocityRelationship}, we find that $\approx93\%$ of the overall pulse speed is attributable to chemotaxis in the case of weak confinement. 

As a final confirmation of this point, we re-run the simulations, but with either growth or chemotaxis removed -- shown by the second and third columns of Fig. \ref{fig:knockouts}, respectively -- thereby isolating the contributions of chemotactic and growth-driven spreading. In the prototypical case of intermediate confinement, both chemotactic and growth-driven spreading play appreciable roles; compare Panels E and H to B in Fig. \ref{fig:knockouts}, as well as the different curves in Figs. \ref{fig:combinedconfinement}B and E. However, in the case of weak confinement, chemotactic spreading dominates, as expected; the simulation without growth (Fig. \ref{fig:knockouts}D) is nearly identical to that incorporating all factors (Fig. \ref{fig:knockouts}A), while the simulation without chemotaxis (Fig. \ref{fig:knockouts}G) yields a population that barely spreads --- also seen by comparing the different curves in Figs. \ref{fig:combinedconfinement}A and D. Therefore, chemotactic propagation dominates under lesser confinement, enabling the population to spread faster as a sharp, coherent pulse.

\subsubsection{Strong confinement} \noindent Population spreading is markedly different in strong confinement. In this case, cells do not form a coherent pulse at all; instead, they continually grow outward as a jammed front (Fig. \ref{fig:knockouts}C), unlike the case of intermediate confinement (Fig. \ref{fig:knockouts}B). Notably, this front does \textit{not} have a well-defined speed at long times, in stark contrast to the cases of weaker confinement explored previously. Instead, the leading edge position progresses as $x_{l}\sim t^\nu$ with $\nu\approx0.5$ at long times, as shown by the solid curve in Fig. \ref{fig:combinedconfinement}C---and thus, the population spreads less effectively. This diffusive scaling of  $x_{l}$ is at odds with the prediction of the classic Fisher–KPP model, commonly used to describe growth-driven spreading, that the population spreads at a constant speed as a traveling wave \cite{cremer2019chemotaxis,narla2021traveling}. Instead, our finding is consistent with the results of agent-based simulations of a growing population of jammed, incompressible cells \cite{farrell2013mechanically}, which also found $\nu\approx0.5$ in the limit of fast nutrient consumption. In this case, front propagation \textit{via} growth of the jammed population lags behind nutrient diffusion --- leading to the diffusive scaling of $x_{l}$ observed in our simulations as well as those of \cite{farrell2013mechanically}. This difference with the prediction of the classic Fisher–KPP model suggests that the logistic form of growth used therein does not adequately describe jammed growth spreading.

This dominant role of growth in driving spreading in the case of strong confinement is expected from the smaller value of $\Lambda$; it is also corroborated by quantification of the different fluxes driving spreading (Fig. \ref{fig:combinedconfinement}F). Removing growth or chemotaxis from the simulation provides a final confirmation of this point; the simulation without growth (Fig. \ref{fig:knockouts}F) yields a population that barely spreads, while that without chemotaxis (Fig. \ref{fig:knockouts}I) is nearly identical to that incorporating all factors (Fig. \ref{fig:knockouts}C) --- also seen by comparing the different curves in Figs. \ref{fig:combinedconfinement}C and F. Hence, growth-driven spreading dominates under stronger confinement, enabling the population to spread diffusively as a jammed front.

\section{Discussion}\label{Discussion}
\noindent Ever since the discovery of bacteria over 300 years ago, lab studies of their spreading have typically focused on cells in unconfined environments such as in liquid cultures or near flat surfaces. However, in many real-world settings, bacteria must navigate complex and highly-confining environments. Thus, motivated by experimental observations of bacterial motility \cite{cisneros2006reversal,drescher2011fluid,croze2011migration,tapasoftmatter,051bhattacharjeenatcomm,bhattacharjee2021chemotactic} and growth \cite{dell2018growing,volfson2008biomechanical} in confined settings, in this paper, we have presented an extended version of the classic Keller-Segel model that incorporates the influence of confinement on bacterial spreading. 

In particular, our extended model treats cellular collisions with rigid surrounding obstacles, cellular collisions with each other, and growth-mediated spreading of jammed populations of cells. As such, it helps to bridge the classic Keller-Segel model of chemotactic spreading -- which does not treat these effects and is therefore only appropriate to describe the spreading of dilute populations in unconfined settings -- and models of growth-driven spreading (e.g., \cite{farrell2013mechanically}) -- which do not treat motility-based spreading and are therefore only appropriate to describe the spreading of highly-concentrated/confined and non-motile populations. Indeed, non-dimensionalizing our extended model reveals the parameter $\Lambda$ that quantifies the confinement-mediated transition between chemotactic spreading (in weak confinement with $\Lambda\gtrsim1$) and growth-driven spreading (in stronger confinement with $\Lambda<1$). Our mathematical analysis also provides a straightforward way to estimate, in general, the relative contributions of chemotaxis and growth to the speed with which a population spreads. 

Furthermore, numerical simulations of the model enable us to examine the implications of this transition for the full dynamics of bacterial spreading. As expected, in weak confinement, a dense inoculum of bacteria rapidly depletes nutrient locally, causing a coherent pulse of cells to detach and continually propagate outward \textit{via} chemotaxis --- as predicted by the classic Keller-Segel model \cite{keller1971traveling,odell1976traveling,keller1975necessary,lauffenburger1991quantitative,fu2018spatial,cremer2019chemotaxis,seyrich2019traveling}. However, with increasing confinement, cellular crowding increasingly hinders both the initial formation of this pulse as well as its long-time propagation speed. Moreover, with increasing confinement, growth plays an increasingly dominant role in driving population spreading --- eventually leading to a transition from fast chemotactic spreading to slow, growth-driven spreading of a jammed front \cite{farrell2013mechanically}. Therefore, confinement is a key regulator of population spreading. 

While chemotactic pulse propagation is well-characterized in unconfined settings \cite{adler1966effect,adler1966science,saragosti2011directional,fu2018spatial}, and conversely, jammed growth expansion has been investigated in some highly-confined settings \cite{dell2018growing,volfson2008biomechanical}, the interplay between these two behaviors has scarcely been studied. Hence, we anticipate that our numerical characterization of this confinement-mediated transition from chemotactic- to growth-driven spreading will help guide future experimental investigations of confined populations. Moreover, because our model describes migration over large length and time scales, we expect it could help more accurately describe the spreading dynamics of bacteria in processes ranging from infections, drug delivery, agriculture, and bioremediation. 

Our extended model represents a first step toward capturing all the biophysical processes underlying these complex dynamics, and necessarily involves some simplifying assumptions and approximations. For example, based on recent experiments \cite{bhattacharjee2021chemotactic}, we treated the influence of cell-cell collisions using a mean-field approach in which the transport parameters $D_{b0}$ and $\chi_0$ are truncated in a cell density-dependent manner; incorporating more sophisticated collective dynamics \cite{dunkel2013fluid,gachelin2014collective,sokolov2012physical,colin2019chemotactic} will be an important extension of our work. Similarly, we described jammed growth expansion by treating the population as an incompressible ``fluid", similar to other models of growing populations \cite{farrell2013mechanically,klapper2002finger,head2013linear} and motivated by some experiments \cite{dell2018growing,volfson2008biomechanical}; developing a more detailed treatment of these dynamics, such as by incorporating cellular deformations \cite{farrell2013mechanically} and possible changes in cellular behavior that may result \cite{chu2018self}, will be a useful direction for future work. Furthermore, a simplifying assumption made in our model is that the solid obstacles that induce confinement are rigid and immovable; incorporating deformations of the surrounding medium will likely give rise to even more complex dynamics that will be interesting to study. Finally, while our model assumes that nutrient diffusion is unimpeded by the solid medium---which is likely to be the case in highly-permeable media such as biological gels and microporous clays/soils---incorporating hindered nutrient diffusion that may arise in other media will likely result in more complex dynamics that future extensions of our work could explore. \\

\section*{Author contributions} \noindent D.B.A. and S.S.D. developed the theory, assisted by discussions with T.B.; D.B.A., J.A.O., and S.S.D. designed the numerical simulations; D.B.A. performed all numerical simulations with assistance from J.A.O.; D.B.A. and S.S.D. analyzed
the data; S.S.D. designed and supervised the overall project. D.B.A. and S.S.D. discussed the results and implications and wrote
the manuscript.

\nolinenumbers

%
%
%






\providecommand{\noopsort}[1]{}\providecommand{\singleletter}[1]{#1}%

\newpage\section*{Supporting Information}
\setcounter{figure}{0}
\makeatletter 
\renewcommand{\thefigure}{S\@arabic\c@figure}
\makeatother

\begin{figure}[h]
\centering
\includegraphics[width=0.7\textwidth]{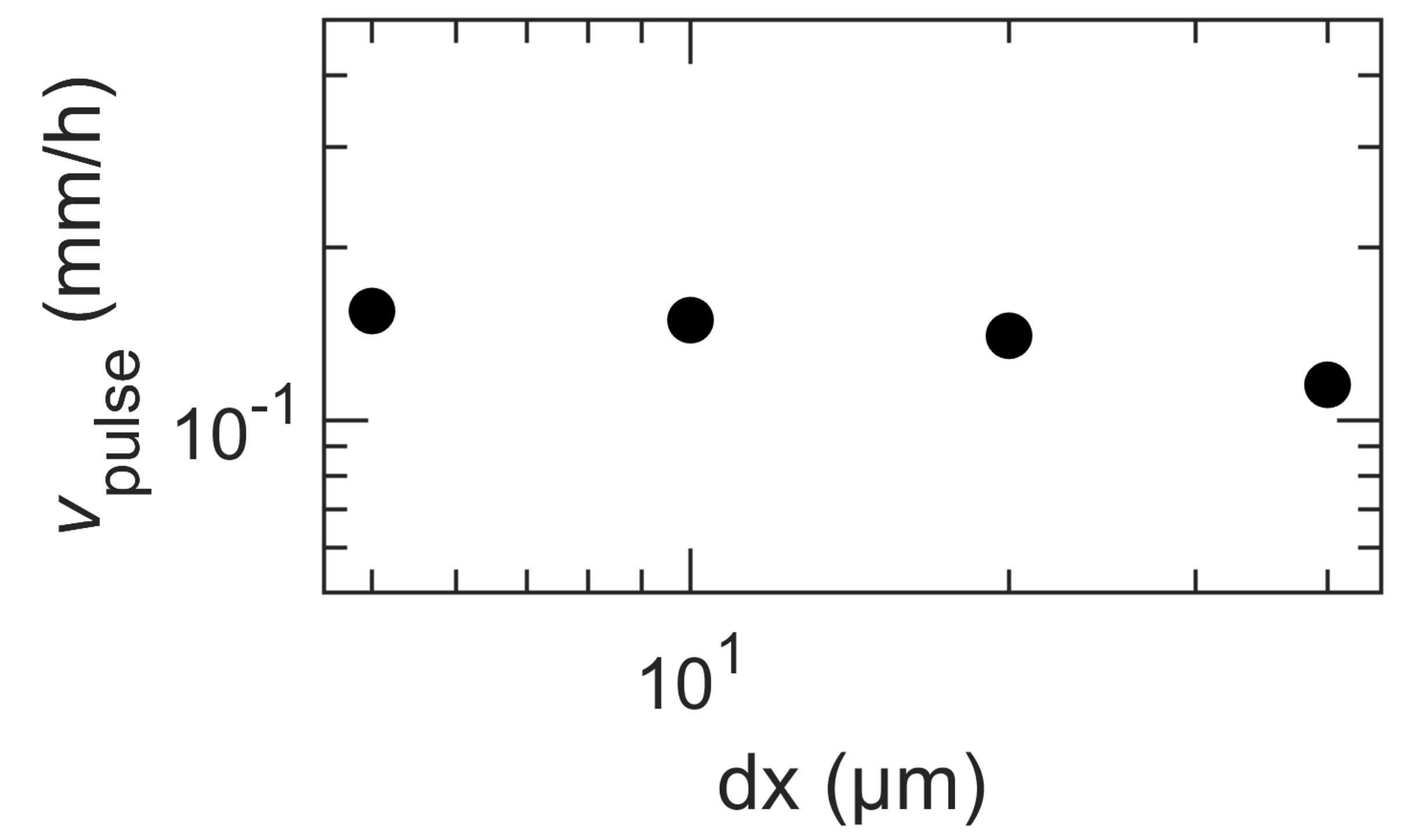}
\caption{To assess the sensitivity of our results to numerical discretization, we repeat the simulation shown in Fig.~\ref{fig:profiles_dense}, which has spatial resolution of $dx = 10$ \textmu{m}, with varying values of $dx$; the time step $dt$ is correspondingly varied as $dt$ = 0.01 s $\times(dx/10$ \textmu{m}$)^2$. As shown in the figure, the final pulse velocity $v_{\rm{pulse}}$ obtained from the simulations is not strongly sensitive to
the choice of numerical discretization.}
\label{fig:converge}

\end{figure}

\subsection*{S1: Determining parameters from experimental data}
\begin{figure}[!h]
\centering
    \includegraphics[width=.8\textwidth]{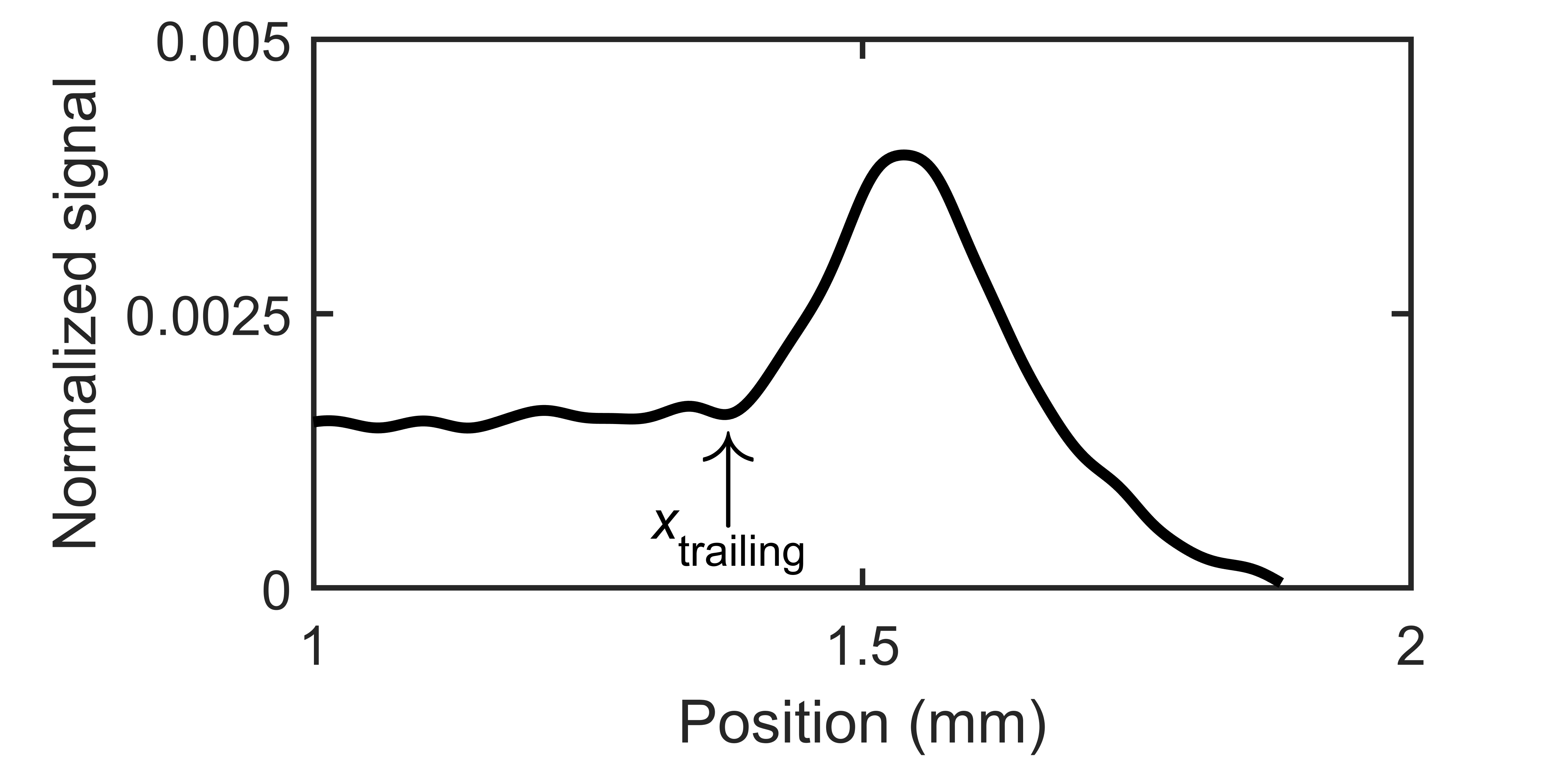}
\caption{Experimental cellular signal of traveling front under intermediate confinement from \cite{bhattacharjee2021chemotactic}. Experiments begin with a dense packed cylindrical inoculum of \textit{E. coli} embedded within a porous media with mean pore size 1.7 \textmu{m}. A pulse forms and propagates outward; the dataset shows the final time point of 10.75 h. The experiment used confocal microscopy of cells constitutively expressing green fluorescent protein; we take the fluorescence data thereby obtained from the mid-plane of the bacterial cylinder and normalize it by the brightest region of the initial inoculum. This normalized cellular signal is then converted to cell density by multiplying with $b_{\rm{max}}=0.95\times 10^{12}$ cells/mL. Arrow indicates location identified as trailing behind the pulse, $x_{\rm{trailing}}$, and the corresponding cellular density is $b_{\rm{trailing}}=1.5\times 10^{9}$ cells/mL. }
\label{fig:expt_profiles}
\end{figure}

\noindent The parameter values for $\kappa$, $c_{\rm{char}}$, and $\chi_0$ are crucial for fitting to experimentally observed pulse propagation speeds. In our previous work \cite{bhattacharjee2021chemotactic}, we chose values of $\kappa$ and $c_{\rm{char}}$ based on previous measurements \cite{cremer2019chemotaxis,croze2011migration}, and therefore only $\chi_0$ was treated as a fitting parameter. While the previous simulations based on these choices reasonably captured the formation and propagation of bacterial pulses observed in experiments, as well as the experimentally-observed dependence of pulse skewness, height, and speed on confinement, the widths of the simulated pulses differed noticeably from the experiments. Because the goal of this present paper is to more closely investigate pulse shape and dynamics, here we develop a new protocol to determine all three parameters $\kappa$, $c_{\rm{char}}$, and $\chi_0$ from the experimental data. The experiment is summarized in \S \ref{implementation}. The primary dataset we use for fitting (shown in Fig. \ref{fig:expt_profiles} and taken from \cite{bhattacharjee2021chemotactic}) is obtained from a late time experimental profile for cells in intermediate confinement (corresponding to \S \ref{Results:intermediate}). 

\textit{Determining $\kappa$.} At long times, given that the pulse is nearly unchanging in time, we take $\partial c/\partial t=-v_{\rm{pulse}}\partial c/\partial \xi$ and $\partial b/\partial t=-v_{\rm{pulse}}\partial b/\partial \xi$. Applying these definitions in Eqs. \ref{eqnKellerSegelc}-\ref{eqnKellerSegelb} and integrating over all space, with the boundary conditions $b(\xi=\infty)=0$, $c(\xi=\infty)=c_\infty$, $b(\xi=0)={b_{\rm{trailing}}}$, and $c(\xi=0)=0$ yields the steady-state relationship between  nutrient influx into the pulse to the cells being shed at the rear, $\kappa= c_\infty\gamma/b_{\rm{trailing}}$. Here, $c_\infty$ was fixed in the experiments to be 10 mM and $\gamma$ was directly measured to be 0.69 h$^{-1}$. Thus, using the $b_{\rm{trailing}}=1.5\times 10^{9}$ cells/mL directly obtained from the experimental profile, we obtain $\kappa=1.3\times10^{-12}$ mM (cells/mL)$^{-1}$ s$^{-1}$. This value, which we use for all the simulations reported here, is in excellent agreement with previously reported values  \cite{cremer2019chemotaxis,croze2011migration}.

\textit{Determining $c_{\rm{char}}$ and $\chi_0$.} Having obtained $\kappa$, we next use the experimental data to determine $c_{\rm{char}}$ and $\chi_0$. To do so, we first re-run the simulation of \S \ref{Results:intermediate} but with $c_{\rm{char}}$ chosen to be either 1, 5, 10, 50, or 100 \textmu{M} --- values that span the range reported in previous experiments \cite{cremer2019chemotaxis,croze2011migration,fu2018spatial}. For each choice of $c_{\rm{char}}$, we then determine the value of $\chi_0$ for which the simulation $v_{\rm{pulse}}$ best matches the experimental value. Then, having fit $v_{\rm{pulse}}$, we pick the value of $c_{\rm{char}}$ for which the simulated pulse width best matches the experimental data. Together, this iterative procedure yields the unique combination of $\{c_{\rm{char}},\chi_0\}$ that best matches the experimental long-time pulse speed and width. We thereby obtain $c_{\rm{char}}=10$ \textmu{M} and $\chi_0=94$~\textmu{m}$^2$s$^{-1}$ for cells in intermediate confinement. Because $c_{\rm{char}}$ is an intrinsic cellular property, and thus does not depend on confinement, we then use this value of $c_{\rm{char}}$ for other simulations testing weak and strong confinement as well. For each of these, we again obtain $\chi_0$ by fitting the long-time $v_{\rm{pulse}}$ between simulations and the experiments. We obtain $\chi_0=3700$ \textmu{m}$^2$s$^{-1}$ and $\chi_0=16$ \textmu{m}$^2$s$^{-1}$ for weak and strong confinement, respectively. 

\subsection*{Descriptions of supporting videos}

\paragraph*{S1 Video.}
\label{S1_Video}
{\bf Numerical simulations of bacterial spreading from a dense inoculum in intermediate confinement.} Video corresponds to Fig. \ref{fig:profiles_dense}. Top shows the dynamics of the cells while bottom shows the corresponding dynamics of the nutrient, quantified by the normalized density $b/b_{\rm{max}}$ and concentration $c/c_{\infty}$, respectively. Lower inset shows the same nutrient data, but with both axes zoomed out).

\paragraph*{S2 Video.}
\label{S2_Video}
{\bf Numerical simulations of bacterial spreading from a dilute inoculum in intermediate confinement.} Video corresponds to Fig. \ref{fig:profiles_dilute}. Top shows the dynamics of the cells while bottom shows the corresponding dynamics of the nutrient, quantified by the normalized density $b/b_{\rm{max}}$ and concentration $c/c_{\infty}$, respectively. Top inset shows the same cellular data, but with the vertical axis zoomed in. Lower inset shows the same nutrient data, but with both axes zoomed out).

\paragraph*{S3 Video.}
\label{S3_Video}
{\bf Numerical simulations of bacterial spreading in weak, intermediate, and strong confinement, also with growth or chemotaxis removed.} Video corresponds to Fig. \ref{fig:knockouts}, but also includes the nutrient profiles in orange (same vertical scales as in Figs. \ref{fig:profiles_dense} and \ref{fig:profiles_dilute}). Different rows show different degrees of confinement, while different columns show full simulations or simulations with growth or chemotaxis removed. The simulations shown in different rows progress over different durations of time.
\end{document}